\documentclass[twocolumn]{aastex631}
\usepackage{graphicx}
\usepackage{enumerate}
\usepackage{amssymb, amsmath, amsfonts}
\usepackage{gensymb}
\usepackage{mathtools, comment}
\usepackage{hyperref}
\usepackage{ulem}
\usepackage{paralist}
\usepackage{xspace}
\usepackage{xcolor}
\usepackage[normalem]{ulem}

\newcommand{\masyr}{\hbox{mas\,yr$^{-1}$}}

\newcommand{\asyr}{\hbox{$\arcsec$\,yr$^{-1}$}}

\newcommand{\Mjup}{\mbox{$M_{\rm Jup}$}}

\newcommand{\addition}[1]{\textcolor{black}{#1}}

\newcommand{\Hipparcos}{\textsl{Hipparcos}\xspace}
\newcommand{\hipparcos}{\textsl{Hipparcos}\xspace}
\newcommand{\htofversion}{1.0.0\xspace}
\newcommand{\orbitcodename}{\texttt{orvara}\xspace}
\newcommand{\Gaia}{\textsl{Gaia}\xspace}
\newcommand{\gaia}{\textsl{Gaia}\xspace}
\newcommand{\ngrst}{\textsl{NGRST}\xspace}
\newcommand{\codename}{\texttt{htof}\xspace}
\newcommand{\htof}{\texttt{htof}\xspace}

\DeclareMathOperator{\arctantwo}{arctan2}

\accepted{July 7, 2021}

\shorttitle{{\tt htof}}
\shortauthors{Brandt et al.}
\submitjournal{AJ}
\accepted{July 7, 2021}

\begin{document}

\title{{\tt htof}: A new open-source tool for analyzing Hipparcos, Gaia, and future astrometric missions.
}

\correspondingauthor{G.~Mirek Brandt}
\email{gmbrandt@physics.ucsb.edu}


\author[0000-0003-0168-3010]{G.~Mirek Brandt}
\altaffiliation{NSF Graduate Research Fellow}
\affiliation{Department of Physics, University of California, Santa Barbara, Santa Barbara, CA 93106, USA}

\author[0000-0002-7618-6556]{Daniel Michalik}
\altaffiliation{ESA Research Fellow}
\affiliation{European Space Agency (ESA), European Space Research and Technology Centre (ESTEC), Keplerlaan 1, 2201 AZ Noordwijk, The Netherlands}

\author[0000-0003-2630-8073]{Timothy D.~Brandt}
\affiliation{Department of Physics, University of California, Santa Barbara, Santa Barbara, CA 93106, USA}

\author[0000-0002-6845-9702]{Yiting Li}
\affiliation{Department of Physics, University of California, Santa Barbara, Santa Barbara, CA 93106, USA}
 
\author[0000-0001-9823-1445]{Trent J.~Dupuy}
\affiliation{Institute for Astronomy, University of Edinburgh, Royal Observatory, Blackford Hill, Edinburgh, EH9 3HJ, UK}

\author{Yunlin Zeng}
\affiliation{Department of Physics, University of California, Santa Barbara, Santa Barbara, CA 93106, USA}
\affiliation{School of Physics, Georgia Institute of Technology, 837 State Street, Atlanta, Georgia, USA}


\begin{abstract}
We present \codename, an open-source tool for interpreting and fitting the intermediate astrometric data (IAD) from both the 1997 and 2007 reductions of \Hipparcos, the scanning-law of \Gaia, and future missions such as the Nancy Grace Roman Space Telescope (\ngrst). \codename solves for the astrometric parameters of any system for any arbitrary combination of absolute astrometric missions. In preparation for later \Gaia data releases, \codename supports arbitrarily high-order astrometric solutions (e.g. five-, seven-,  nine-parameter fits). Using \codename, we find that the IAD of 6617 sources in \Hipparcos 2007 might have been affected by a data corruption issue. \codename integrates an ad-hoc correction that reconciles the IAD of these sources with their published catalog solutions. We developed \htof to study masses and orbital parameters of sub-stellar companions, and we outline its implementation in one orbit fitting code (\orbitcodename). We use \codename to predict a range of hypothetical additional planets in the $\beta$~Pic system, which could be detected by coupling \ngrst astrometry with \Gaia and \Hipparcos. \codename is \texttt{pip} installable and available at \url{https://github.com/gmbrandt/htof}.
\end{abstract}. 


\section{Introduction}

Astrometric missions like \Hipparcos or \Gaia measure absolute positions, parallaxes, and motions for the stars that they survey \citep{2012PerrymanAstrometryHistory}. These missions observe a star many times and fit a model to that star's motion in the plane of the sky, also known as the star's ephemeris. A star's apparent motion is most simply modeled as parallactic motion (motion induced by the earth's orbit) together with a constant velocity; encoded in five parameters. The \Hipparcos \citep{HIP_TYCHO_ESA_1997, vanLeeuwen_2007} and \Gaia \citep{Gaia_General_2016, Gaia_General_2018, Gaia_Astrometry_2018, 2020GaiaEDR3_catalog_summary} catalogs report the five best-fit astrometric parameters: two position parameters at a reference epoch, two proper motions, and the parallax. The five-parameter astrometric model, however, is only an approximation: even a single star with a constant 3D velocity exhibits apparent accelerations when its motion is projected into spherical coordinates (perspective acceleration; \citealp{Michalik+Lindegren+Hobbs+etal_2014, 2021arXiv210509014L}). Stars may also accelerate due to visible and/or unseen companions, or even due to their orbits through the Galaxy.

The influence of unseen companions may be modeled as a higher-order astrometric fit with seven or nine parameters, or as a full Keplerian orbit. The first full orbital fits \addition{of \hipparcos astrometry} were done by the \hipparcos team themselves \citep{1997A&A...323L..53L}. Work over the following decades built on this work to include brown dwarfs and giant planets (e.g., \citealp{2001ApJ_Zucker_Shay, 2010A&A...509A.103S, 2011A&A...527A.140R, Sahlmann+Segransan+Queloz_2011, Snellen+Brown_2018}). The most recent additions to this field are covering harder to detect and less massive exoplanets, leveraging the long-term proper motion anomalies between \hipparcos and \gaia \citep{brandt_cross_cal_gaia_2018, Brandt_Dupuy_Bowler_2018, Calissendorff+Janson_2018, Dupuy+Brandt+Kratter+etal_2019, 2019MNRAS.490.1120F, 2019AA_Kervella, 2019MNRAS_Fabo_Feng_etal_eps_indiA, 2020AA_Maire_HD19467, 2020ApJ_Currie_Thayne_HD33632, Brandt2020betapicbc}.

For companions with orbital periods $\lesssim$10\,yr, proper motion accelerations within \Hipparcos and \Gaia become important to model. The intermediate astrometric data (hereafter, IAD) of the \Hipparcos mission---the individual position measurements and uncertainties for every star---are available. The \Gaia IAD for every star will be unavailable for many years, but the spacecraft's scanning law---its predicted scan angles and observational epochs as a function of sky position---is available through the \Gaia Observation Forecast Tool \citep[GOST;][]{gaia_gost_user_manual}.\footnote{\url{https://gaia.esac.esa.int/gost/}}. Several groups have used GOST predictions to forward model \gaia astrometry in the context of orbit fitting (e.g., \citealp{brandt_gliese_229b_mass_htof, Nielson+DeRosa+etal+betapicc2019, 2020A&A...640A..73D})

In this paper, we present a tool, \codename \citep{htof_zenodo}, that uses the IAD for \Hipparcos and the scanning law of \Gaia to fit sky paths. Although \codename extends beyond the hundred thousand sources of \Hipparcos, \codename stands for the Hundred Thousand Object Fitter in homage to the hundred-thousand-proper-motion project \citep{htpm_JHJ_2012AA, Michalik+Lindegren+Hobbs+etal_2014}. For both \Gaia and \Hipparcos, \codename can compute fits with arbitrarily many parameters, e.g., five-, seven-, and nine-parameters. Given the \Hipparcos IAD for a star, \codename computes likelihoods and parameter errors directly comparable to the catalog values for that star. For \Gaia, \codename is currently limited to approximate likelihoods and errors because the actual observations and their uncertainties are not yet available. 

\codename's capabilities make it well suited for exploring many problems such as simulating the planets which could be discovered by future astrometric missions. The likelihoods computed by \codename also enable full Keplerian orbital fits if combined with common-reference-frame and cross-calibrated positions and errors from \Hipparcos and \Gaia, e.g. from the \Hipparcos-\Gaia Catalog of Accelerations \citep{brandt_cross_cal_gaia_2018}. \codename was used in a number of papers exploiting the latter technique for studying dynamical masses and orbital parameters of sub-stellar companions in nearby stellar systems \citep{brandt_gliese_229b_mass_htof, Brandt2020betapicbc, 2021AJ....161..106B, Bowler2021accepted}.

This paper is structured as follows. In Section \ref{sec:fitting_and_parsing_astrometry}, we describe \Gaia, \Hipparcos, fitting astrometry, and parsing the IAD. In Section \ref{sec:HTOF}, we discuss \codename along with its implementation, validation, and limitations. 

In Section \ref{sec:hipparcos_2_fixes}, we discuss a suspected data corruption affecting 6617 sources in the \hipparcos 2007 IAD. We have developed an ad-hoc correction algorithm to recover their best-fit residuals. It is readily integrated in \codename.
We then discuss two major use cases. We outline \codename's utility in orbit fitting in Section \ref{sec:hipparcos_only_application}. In Section \ref{sec:mission_forecasting}, we demonstrate how \codename can predict what new planets could be discovered by combining \Gaia astrometry with measurements from the Nancy Grace Roman Space Telescope (\ngrst) \citep{wfirst_spergel_2015}. We summarize our results in Section \ref{sec:conclusions}.

\section{Observing and fitting an astrometric sky path}\label{sec:fitting_and_parsing_astrometry}

\Gaia and \Hipparcos observe stars by scanning the sky in a repetitive pattern, called the scanning law \citep{1998AAS130157V, Gaia_General_2016}. For both satellites, the continuous scans along overlapping great circles mean that each observation is precise in the direction along the scan (along-scan; AL), and imprecise perpendicular to the scan (across-scan; AC). This renders each observation effectively one-dimensional \citep{Gaia_General_2016}. 

\begin{figure}
    \centering
    \includegraphics[width=0.4\textwidth]{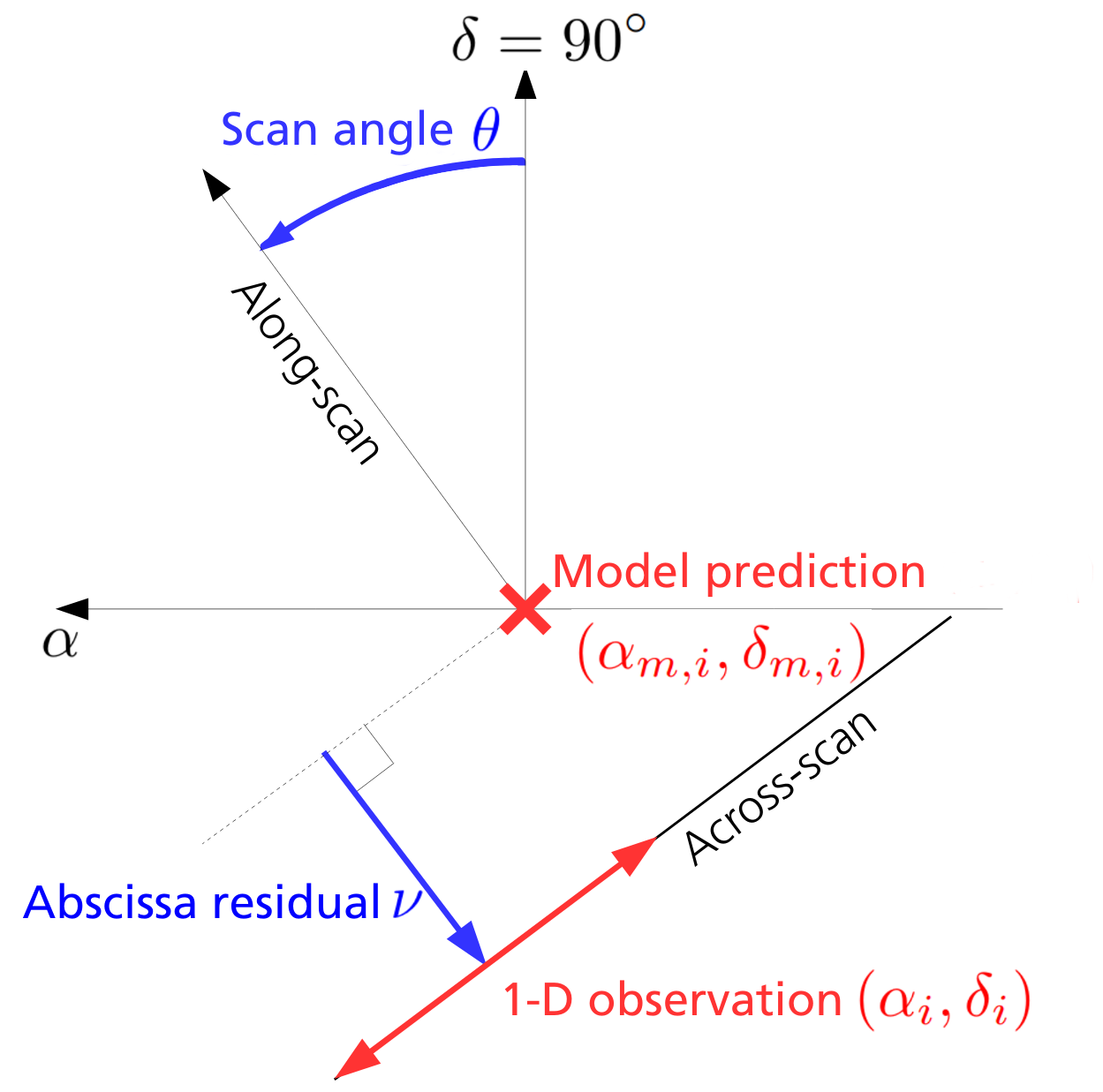}
    \caption{Definition of the scan angle $\theta$ and the abscissa residual $\nu$ between the one-dimensional observation ($\alpha_i, \delta_i$) and the best-fit astrometric model ($\alpha_{m, i}, \delta_{m, i}$).}
    \label{fig:abscissa-theta}
\end{figure}

Figure \ref{fig:abscissa-theta} shows the definitions we use for the scan-angle $\theta$ of the observation and the abscissa residual, consistent with the definitions given in  \cite{Gaia_Solar_System_2018} and online.\footnote{\url{https://www.cosmos.esa.int/web/gaia/scanning-law-pointings}} The scan angle $\psi$ of the 2007 \hipparcos re-reduction is defined differently than for \gaia, but this is handled by a conversion discussed in Section \ref{sssec:hip2}. In Figure~\ref{fig:abscissa-theta}, the spacecraft spin direction is indicated by the AL arrow, the spin axis is perpendicular to it in the AC direction. The (essentially one-dimensional) error-ellipse of an observation is indicated by the red double-headed arrow. \Hipparcos and \Gaia fit sky paths to the abscissa residuals; the model sky paths are collectively referred to as \textit{the astrometric solution}. The shortest distance between the prediction of the best-fit astrometric model at the observational epoch and the 1-d observed position is called the abscissa residual $\nu$. 

The astrometric solution for a given star is the set of model parameters that minimizes the sum of the squared AL abscissa residuals weighted by the inverse of the AL variances. A good astrometric solution requires data with good geometric and temporal coverage, i.e., many different scan angles at many different epochs. In the following subsections, we discuss how to obtain the best-fit astrometric parameters from the IAD.

\subsection{Fitting the Sky Path}\label{ssec:fitting_astrometry}

For a five-parameter astrometric model, the observed position on the sky is the sum of the parallax, proper motion, and position multiplied by known functions of time. Higher order models include acceleration and jerk terms. Fitting the ephemerides under 5, 7, or 9-parameter models is a linear problem, and may be solved by minimizing the chi-squared of the abscissae with respect to the astrometric parameters. The chi-squared minimization problem can be transformed into a matrix equation which can then be solved by any standard linear algebra method, e.g., singular value decomposition in \codename. The model for the star's motion across the sky that we adopt is
\begin{align}{}\label{eq:model_parallax}
\begin{aligned}
    \alpha_{m,i} &= \varpi f_\varpi[t_i] + \sum_{n=0}^N \frac{a_n}{n!} (t_i - t_{\rm ref}) ^n \\
    \delta_{m,i} &= \varpi g_\varpi[t_i] + \sum_{n=0}^N \frac{b_n}{n!} (t_i - t_{\rm ref})^n
\end{aligned}{}
\end{align}
where $(\delta_{m,i}, \alpha_{m,i})$ is the predicted model declination and right-ascension at time $t_i$. $a_n, b_n$ are the astrometric parameters excluding parallax (e.g.~$b_1 = \mu_{\delta}$), $N$ is the degree of the fit ($N=1$ for a five-parameter fit), $\varpi$ is the parallax angle, and $f_\varpi[t_i], g_\varpi[t_i]$ are the parallax factors \citep{hipparcos_rereduction_book}. These are the perturbations in declination and right-ascension, respectively, due to parallax alone for a parallax angle of unity. $f_\varpi[t_i] \cos(\theta_{\rm scan}) + g_\varpi[t_i] \sin(\theta_{\rm scan})$ is equal to $\frac{\partial r_i}{\partial \varpi}$, the derivative of the residual with respect to parallax. We use modules from \texttt{astrometric-sky-path}\footnote{\url{https://github.com/agabrown/astrometric-sky-path}} to compute the parallax factors. 

The \hipparcos catalog abscissa residuals $\nu_i$ are with respect to the along-scan direction only (for both the 1997 and 2007 catalogs). The matrix equation can be constructed, and in turn the best-fit astrometric parameters can be solved for, purely in terms of the along-scan derivatives and abscissa residuals \citep{HIP_TYCHO_ESA_1997}. Alternatively, one can first cast the measurements into the equatorial coordinate system and construct the matrix equation from those right-ascension and declination observations. The former approach is closer to how the \hipparcos team handled the data \citep{HIP_TYCHO_ESA_1997}, while \codename uses the latter approach, whereby we minimize Equation \eqref{eq:chi_squared_fit}. The two are equivalent, but we employ the latter because then the input data are more familiar: right ascension and declination instead of along-scan abscissa residuals and derivatives.

We therefore assume that we have: the covariance matrix $\mathcal{C}_i$ for an observation $i$ which occurred at epoch $t_i$ and the corresponding right-ascension $\alpha_i$ and declination $\delta_i$. We obtain the $\delta_i, \alpha_i$ by decomposing each abscissa residual $\nu_i$ into the two equatorial components. We find the best-fit astrometric parameters by minimizing 
\begin{align}\label{eq:chi_squared_fit}
    \chi^2 = \sum_i \left(\begin{matrix}\alpha_{m,i} - \alpha_{i}&, \, \delta_{m,i} - \delta_{i}\end{matrix}\right) \mathcal{C}_i^{-1} \left(\begin{matrix}\alpha_{m,i} - \alpha_{i}\\\delta_{m,i} - \delta_{i}\end{matrix}\right),
\end{align}

In detail, the parallactic motion of any object with a non-zero proper motion will change as it moves through the sky. The parallax factors are functions of position as well as time. Nonlinear effects such as perspective acceleration are negligible for most stars. For example, \cite{Gaia_Astrometry_2018} lists only 53 stars where nonlinear effects warrant correction, one of which is Barnard's star ($\mu \approx 10$\asyr; \citealp{1935AJ.....44...74V, Luyten-high-proper}). We therefore always assume that the model (Equation \ref{eq:model_parallax}) is linear.

We solve for the best-fit parameters by taking the partial derivative of Equation \eqref{eq:chi_squared_fit} with respect to each of the free parameters. The partials yield a matrix equation of full rank (e.g., a $9 \times 9$ matrix if $N = 3$ in Eq.~\ref{eq:chi_squared_fit}) which is linear in every parameter. We solve for the best-fit parameters with singular value decomposition.

\subsection{Obtaining astrometry from \Hipparcos IAD and the \gaia scanning law}\label{ssec:parsing_astrometry}
To recover a solution for each source we need the epochs of observation, the source's position in the sky at each epoch, and the covariance matrix of the observation at each epoch. In this section, we detail how to obtain and parse those quantities from \hipparcos and \gaia.

For both \hipparcos reductions, the IAD are enough to calculate full positions and covariance matrices. For \Gaia, we use the predicted scanning law as a proxy for the IAD. We calculate positions assuming the Gaia catalog astrometric parameters, and calculate the epochs and covariance matrices for each observation.

Table~\ref{table:angle_and_epoch} summarizes how to extract the scan angles, observation times and along-scan errors from the IAD from both \Hipparcos reductions and the \Gaia scanning law. Each variable and column in Table~\ref{table:angle_and_epoch} is named in accordance with each astrometry product. We now turn to details of the IAD and the scanning law.

\begin{deluxetable*}{cccccc}
\tablewidth{0pt}
    \tablecaption{Summary of extracting the scan angles, epochs, residuals, and along-scan errors ($\sigma_{\rm AL}$) from the three catalogs.\label{table:angle_and_epoch}}
    \tablehead{
    \colhead{Reduction} & \colhead{Scan angle (radian)} & \colhead{Epoch} & \colhead{$\sigma_{\rm AL}$} & \colhead{AL residuals} & \colhead{Reference\tablenotemark{1}}}
    \startdata
    Hip 1     & $\arctantwo \left(\texttt{IA3}, \texttt{IA4} \right) $  & $ J1991.25 + \frac{\texttt{IA6}}{\texttt{IA3}}$ OR $J1991.25 + \frac{\texttt{IA7}}{\texttt{IA4}}$  & \texttt{IA9} & \texttt{IA8} & vLE98 \\
    Hip 2 DVD     & $\arctantwo \left( \text{Column 4}, \text{Column 5} \right) $  & Column 2 $+ \, J1991.25$ & Column 7 & Column 6 & vL07 \\
    Gaia DR2  &   \texttt{scanAngle} & \texttt{ObservationTimeAtBarycentre} &   \nodata & \nodata & FHJ19 
    \enddata
        \tablecomments{Hip 1 refers to the 1997 reduction \citep{HIP_TYCHO_ESA_1997}, and Hip 2 to the 2007 re-reduction \citep{vanLeeuwen_2007}.}
        \tablenotetext{1}{References abbreviated as vLE98 \citep{1998AAS130157V}, vL07 \citep[][Table G.8]{hipparcos_rereduction_book}, and FHJ19 \citep{gaia_gost_user_manual}.}
\end{deluxetable*}

\subsubsection{The original (1997) Hipparcos catalog}

The IAD for the original \Hipparcos catalog \citep{HIP_TYCHO_ESA_1997} are available from the Strasbourg Astronomical Data Center (CDS) FTP archive.\footnote{\url{http://cdsarc.u-strasbg.fr/ftp/I/239/version_cd/cats/hip_i.dat.gz}} The IAD give the scan angles, abscissa residuals \addition{relative to} the first five parameters of the best-fit sky path, formal AL errors for each observation, and time of observation. Each file from the IAD has ten columns, referred to in the documentation as \texttt{IA1} through \texttt{IA10}. In the second column (\texttt{IA2}), each row is labelled either \texttt{N} or \texttt{F} for NDAC for FAST, depending on which consortium reduced the data. The formal AL errors ($\sigma_{AL}$) are in column \texttt{IA9} and the abscissa residuals are given in column \texttt{IA8}. The scan angle is $\arctantwo \left( \texttt{IA3}, \texttt{IA4} \right)$ \citep{1998AAS130157V}.

The astrometric parameters in the original \Hipparcos catalog are those which best-fit the merged IAD. We merge the NDAC and FAST IAD following the procedure described by the \Hipparcos team (see Section 3.2 of \citealt{1998AAS130157V}). For each epoch that contains both NDAC and FAST data, we average the values from both consortia weighted by Best-Linear-Unbiased-Estimator (BLUE) weights. These weights, a pair per epoch, are determined uniquely by the covariance matrix constructed according to Equation 7.11 on page 377 of Volume 3 of \cite{HIP_TYCHO_ESA_1997}, using the NDAC-FAST correlation coefficient in column \texttt{IA10}. We exclude observations rejected from the \Hipparcos merged solution \citep{HIP_TYCHO_ESA_1997}, which are designated by a lowercase letter in the  consortium designation.

\subsubsection{The 2007 re-reduction of the Hipparcos data}\label{sssec:hip2}

Ten years after the publication of the original \Hipparcos catalog, \cite{vanLeeuwen_2007} re-reduced the raw data to produce new astrometric fits.  
At the time of writing, there are three sets of the \Hipparcos re-reduction catalog. First, and oldest, is the Digital Versatile Disc (DVD) included with \cite{vanLeeuwen_2007}, which contains the first published version of the catalog as well as the IAD used to construct it. Second is the Vizier catalog, \footnote{\url{http://vizier.u-strasbg.fr/viz-bin/VizieR?-source=I/311}} which does not include the IAD. Third, and most recent, is the data distributed with a Java tool. \footnote{\url{https://www.cosmos.esa.int/web/hipparcos/interactive-data-access}} This tool includes the IAD but not in per-source human readable data files as are on the DVD. We have extracted the IAD from the Java tool \citep[private comm.]{floor_private_comm}; \addition{a copy of the dataset is available at \url{https://www.cosmos.esa.int/web/hipparcos/hipparcos-2}}. All three sources give data which are similar for every star, yet not identical. One must consistently use the IAD together with the catalog it accompanied. For example, one should not mix catalog values from Vizier with the IAD from the DVD. For most sources, results using the IAD from the DVD are indistinguishable from results using the IAD from the Java tool. One advantage of the Java tool data is that rejected transits are clearly denoted by negative AL errors. Rejected observations are not designated in the DVD IAD.

The columns of the residual records of the 2007 IAD (both the DVD and Java tool) are defined in Table G.8 of \cite{hipparcos_rereduction_book}. The fourth column gives $\cos(\Psi) = \sin(\theta)$, and the fifth column gives $\sin(\Psi) = \cos(\theta)$, where $\theta$ is the scan angle and $\Psi = \pi/2 - \theta$. We index the first column as 1 to be consistent with the original reduction's documentation. The AL errors are given in column 7 for the re-reduction. The abscissa residuals are in column 6.
\addition{For sources with 7 and 9 parameter solutions, the residuals in the \hipparcos 2007 IAD are relative to the full 7 or 9 parameter solutions. This is in contrast to \hipparcos 1997 where the residuals are defined with respect to the skypath described by only the first five parameters, regardless of solution type.}

\subsubsection{The Gaia Observation Forecast Tool (GOST)}\label{sssec:gost_tool}
\Gaia Data Release 1 (DR1), DR2, and EDR3 \citep{Gaia_General_2016,Gaia_General_2018,Gaia_Astrometry_2018, 2020GaiaEDR3_catalog_summary} contain model parameters (positions, proper motions, parallaxes, and their uncertainties and covariances), but neither the epoch astrometry, the along-scan residuals, nor the per-epoch covariance matrices. The full epoch astrometry will be released in the final catalog of the nominal mission.\footnote{\url{https://www.cosmos.esa.int/web/Gaia/release}\label{gaia_release_schedule_esacosmos}} \codename uses GOST as a stand-in for these data. GOST provides predicted observations and scan angles for any \gaia source. The GOST data should be downloaded manually as a \texttt{.csv} (see the \texttt{readme} with the \codename source code), but a future version of \codename may automate this. \codename places the GOST data into a suitable form for fitting as if they were the \Gaia IAD. GOST-predicted scans, parsed by \codename, are used in a variety of orbital analyses of brown dwarf companions to help constrain companion masses and orbital inclinations. For instance \citet{brandt_gliese_229b_mass_htof, Brandt2020betapicbc, 2021AJ....161..106B, Bowler2021accepted}. However, there are important details to consider with the planned observations.

Not all of the planned observations will be used in the astrometric solution. Some predicted scans will represent missed observations (satellite dead times), executed but unusable observations (e.g.,~from cool-down after decontamination), or observations rejected as astrometric outliers.  Rejected observations could be corrupted due to, e.g.,~micro-clanks, scattered light from a nearby bright source, crowded fields, micro-meteoroid hits, etc.\footnote{\url{https://www.cosmos.esa.int/web/gaia/dr2-data-gaps}\label{dr2_deadtimes_footnote}} Such problematic observations do not constrain the \Gaia astrometric solution. In practice, up to 20\% of \Gaia observations predicted by GOST are excluded from the analysis that produced DR2 \citep{gaia_scanning_law_info}. The largest stretches of dead times and rejected observations are published as astrometric gaps. \codename accounts for the published astrometric gaps for DR2.\footnote{See footnote \ref{dr2_deadtimes_footnote}} and EDR3 (Table 1 of \citealp{Lindegren+Klioner+Hernandez+etal_2020}).

\subsection{Constructing the covariance matrix for each observation}\label{ssec:covariance_matrix}
We need the inverse covariance matrix for every observation, $\mathcal{C}^{-1}_i$, to evaluate and minimize $\chi^2$ (Equation \ref{eq:chi_squared_fit}). We construct $\mathcal{C}^{-1}_i$ from the scanning law or the IAD. We define the ratio of the AC to AL variances as $r$ and assume these two measurements to be independent.
By default, \codename has $r = \infty$ because \hipparcos and \gaia both optimize for AL. For \hipparcos, we have the individual AL uncertainties for all observations. For \gaia, we do not and so we assume the AL error to be the same for all epochs of a given source. The covariance matrix at each epoch is then correct up to a (unknown) constant (the along-scan error). \footnote{See section 6.1.1 of \citet{2021_six_masses_GBrandt} for a discussion of the limitation of the uniform AL error assumption in the context of Gl~229.}

Given a scan angle $\theta_i$, $r$, and AL error $\sigma_{AL, i}$, for the $i^{\rm th}$ observation, we construct the covariance matrix $\mathcal{C}_i$ by rotating the covariance matrix in the scan basis into the equatorial coordinate system as follows.
\begin{align}\label{eq:covariance_matrix}
    \mathcal{C}_i &= \sigma^2_{AL, i} \, \, R \left(\theta_i \right) \left(\begin{matrix}1&0\\0&r\end{matrix}\right) R^T \left(\theta_i \right) \\
    \text{where } & R(\theta_i) = \left(\begin{matrix} 
    \sin \theta_i & - \cos \theta_i\\ \cos \theta_i & \sin \theta_i 
    \end{matrix}\right) \nonumber
\end{align}

In Eq.~\eqref{eq:covariance_matrix}, $R(\theta)$ is the inverse of the transformation matrix in Eq.~(3) of \cite{Gaia_Solar_System_2018} and $R^T(\theta)$ its transpose. For \Gaia, we assume uniform AL errors and set $\sigma^2_{AL}$ to unity for every observation of a given source. We multiply each \Hipparcos re-reduction $\sigma_{AL}$ by the scaling factor $u$ in Equation B.4. of \cite{Michalik+Lindegren+Hobbs+etal_2014}. This is a renormalization to match \citet{hipparcos_rereduction_book}, where the errors were normalized such that the reduced $\chi^2$ of the fit was unity. The inverse covariance matrix is 
\begin{align}\label{eq:inverse_covariance_matrix}
    \mathcal{C}_i^{-1} &= \frac{1}{\sigma^2_{AL, i}} \, \, R \left(\theta_i \right) \left(\begin{matrix}1&0\\0&1/r\end{matrix}\right) R^T \left(\theta_i \right)
\end{align}
where $1/r = 0$ for \Hipparcos and by default in \codename. With the inverse covariance matrix for each observed right-ascension $\alpha_i$ and declination $\delta_i$, we generate the astrometric model which best fits those observations according to Section \ref{ssec:fitting_astrometry}.

\section{\codename}\label{sec:HTOF}
\codename is an open-source \textit{Python} package available at \url{https://github.com/gmbrandt/htof} which parses and fits astrometric data according to the methods discussed in Section \ref{sec:fitting_and_parsing_astrometry}.

\subsection{User guide and core features}\label{ssec:usingHTOF}

\codename is designed to fit 5-, 7-, and 9-parameter models to any \hipparcos or \gaia source. It can calculate the full covariance matrix of the fit, calculate central epochs of observation, perform an ad-hoc correction of the 2007 IAD for 6617 sources (Section \ref{sec:hipparcos_2_fixes}), and combine astrometric missions (useful for forecasting; Section \ref{sec:mission_forecasting}). It possesses a variety of other minor features. \codename \textit{does not} provide an automated download (as of version \htofversion) of the \hipparcos IAD (either reduction) or \gaia GOST data. These data must be downloaded manually.

\codename can analyze data from arbitrary astrometric missions. \codename is centered around \texttt{DataParser} subclasses which parse the IAD or scanning law into scan angles, errors, etc., and a single \texttt{AstrometricFitter} class which generates an astrometric solution from data within any of the numerous \texttt{DataParser} subclasses. The \texttt{DataParser} classes hold the data in a consistent way so that the single fitter class, \texttt{AstrometricFitter}, can fit an astrometric sky path. There is also a convenience \texttt{Astrometry} class, which wraps all the functionality of the data parser and fitting classes. For examples of fits and other functionality, we refer the reader to the documentation \texttt{readme} and the \textit{Jupyter} example notebook that is packaged with the source code. In the following, we review \codename's core functionality of astrometric fitting.

\texttt{Astrometry.fit(ra, dec)} returns an array of the best-fit astrometric parameters. If the fit degree is 1 and \texttt{Astrometry} is instantiated with \texttt{use\_parallax=True}, this array will be $(\varpi, \alpha_0, \delta_0, \mu_{\alpha*}, \mu_{\delta})$: the parallax angle, the right-ascension and declination (relative to the reference right-ascension and declination provided to \texttt{Astrometry()}), the proper motion in right-ascension, and the proper motion in declination --- by default with units of mas and \masyr. If \texttt{fit(ra, dec, return\_all=True)} is used, then the fit will return the best-fit parameter array, an array of the standard errors of each parameter, and the formal chi-square of the fit. The \texttt{AstrometricFitter} can provide the full covariance matrix if it is desired (not just the errors on each parameter), examples of this are in the Jupyter notebook with the source code.

Fitting requires providing a reference epoch around which to center the observations. For \hipparcos 1997 catalog solutions, the consortia used J1991.25. The choice of reference epoch will affect the values, errors, and covariances for all but the highest-order terms in any fit.  For example, changing the reference epoch for a five-parameter fit will affect the fitted reference positions, their errors and covariances, but not the proper motions. Choosing a reference epoch far away from the optimal epoch (e.g.,~choosing the year 1000 for \hipparcos instead of 1991.25) will result in numerical instability and a poorly constrained solution.

\subsection{Validation of the fits to the Hipparcos IAD}\label{ssec:validation}

\codename can reproduce the catalog solutions for the vast majority of sources in both \Hipparcos reductions from the IAD. At the time of this publication, \codename is designed for sources with 5, 7, or 9-parameter solutions, from both \Hipparcos reductions. Future versions of \codename may incorporate sources with VIM (V), stochastic (X), component (C), or orbital (O) solutions. We point the reader to the  \texttt{readme} with the source code for the up-to-date capabilities of the code.

In the current version (\htofversion), there is a subset of the \Hipparcos five-, seven-, and nine-parameter sources for which the best-fit \codename parameters disagree slightly with the catalog values. For the vast majority of sources, \codename matches the catalog values either exactly after rounding, or with a scatter that is attributable to round-off within the intermediate astrometric data. We found that offsets up to $\pm$0.01~$\masyr$ in both right-ascension or declination could be attributed to round off based on Monte-Carlo simulations of randomly selected disagreeing sources. There are only a small number of sources that \codename cannot fit well: 120 sources disagree more than 0.1~$\masyr$ in proper motion in either direction. We label these 120 sources as ``flagged'' and provide a list of them in the file \texttt{htof/data/hip1\_flagged.txt}. A 0.1~$\masyr$ disagreement is typically less than 1/2 of a standard error for the discrepant sources. The fraction of discrepant sources is similar for the \Hipparcos 2007 re-reduction, but only after applying our proposed ad-hoc correction to the IAD.

\section{Ad-hoc correction to 6617 sources in Hipparcos 2007}\label{sec:hipparcos_2_fixes}

\Hipparcos 2007 was a momentous effort largely by one individual.  It is and remains an invaluable record of stellar astrometry. As with every data set, deeper analysis sometimes reveals new ways of interpreting the data. Here we propose an improvement to the \hipparcos 2007 IAD.

We found 6617 sources from \hipparcos 2007 for which the astrometric solution implied by the IAD is significantly discrepant with the published catalog parameters. Two examples are HIP 651 and 44050. 
Discrepant sources are present both in the Java tool and in the DVD IAD. The subset of discrepant sources is almost, but not exactly, identical between the two. However, we limit our discussion of the issue and our ad-hoc correction to the IAD from the Java tool. The Java tool includes a clear designation of rejected transits. Without knowing the regularly rejected transits, our ad-hoc correction algorithm becomes combinatorially unfeasible. There is ongoing work to understand these 6617 sources in more detail \citep[private comm.]{private_comm_floor_june_8_2021}.

To identify the issue, we first read in the IAD for every \hipparcos source. We compute the partial derivatives of $\chi^2$ with respect to each of the astrometric parameters, normalized by the standard errors to make them dimensionless (e.g., $\sigma_{\mu \alpha}\frac{\partial \chi^2}{\partial \mu_{\alpha}}$). If the residuals from the IAD correspond to a solution that minimizes $\chi^2$, then these partials will all be zero within round-off error. For most sources, the quadrature sums of the scaled partial derivatives are indeed nearly zero, ranging from $10^{-2}$ to $10^{-7}$.
For the 6617 sources that we cannot refit, these partial derivative sums are significantly non-zero (e.g., $>1$) -- implying that the best-fit residuals (data minus model) listed in the IAD are not the residuals with respect to the model that minimizes $\chi^2$. As another check, we also refit the IAD with the appropriate astrometric model (5, 7, or 9-parameters) and compare the refit proper motions to those in the catalog. The refit proper motions do not match the catalog values for the discrepant sources.

\begin{figure}[!ht]
    \centering
    \includegraphics[width=\linewidth]{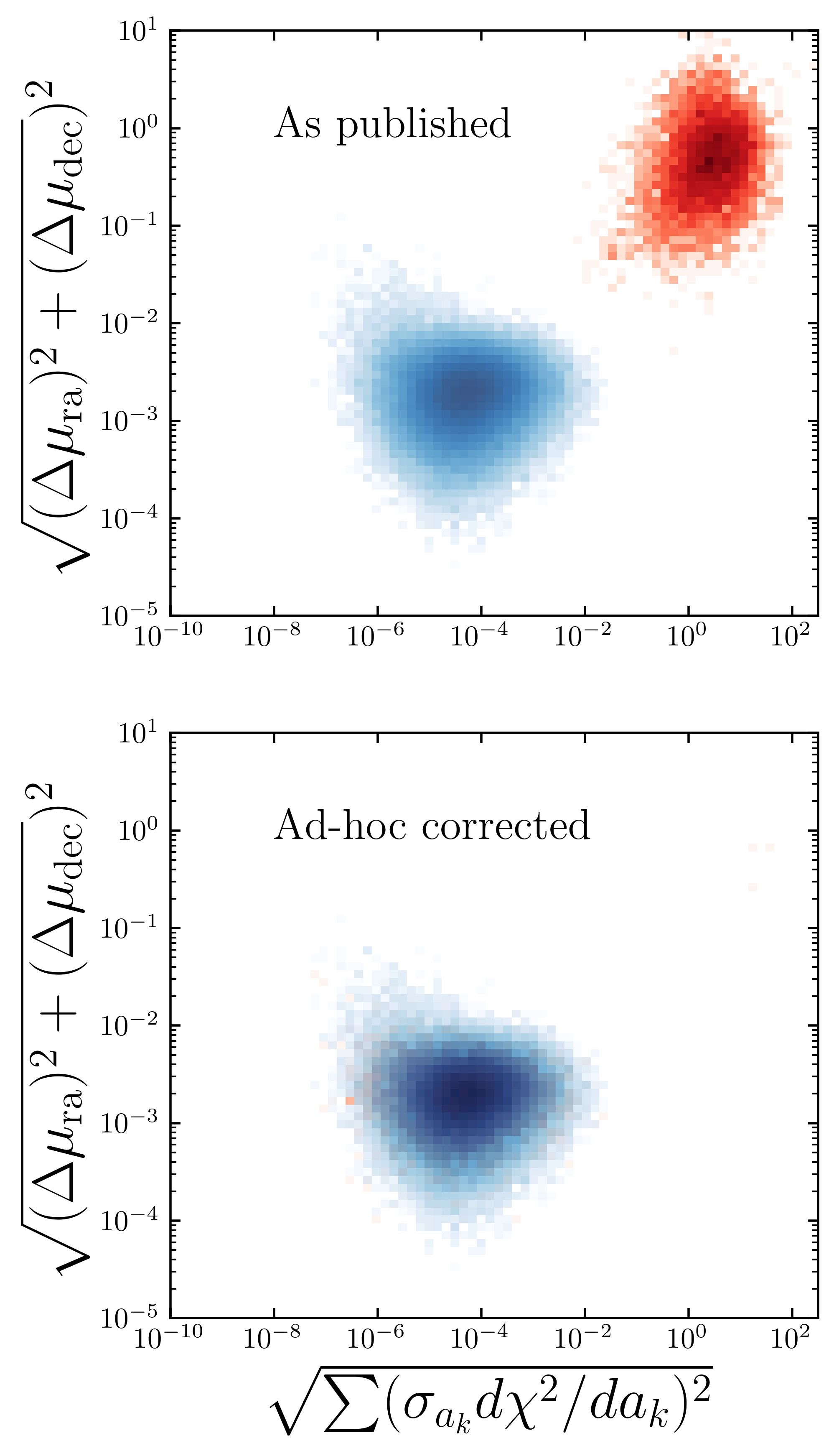}
    \caption{Distributions of the proper motion difference and quadrature sums of partial derivatives of fits to the \hipparcos Java tool IAD before (top panel) and after (bottom panel) the \codename IAD ad-hoc correction. All 115,100 five-parameter sources are shown. The vertical axis shows the quadrature sum of the proper motion  discrepancies (difference between the catalog values and the values obtained by refitting the IAD). The horizontal axis shows the five partial derivatives of chi-squared ($\chi^2$) also added in quadrature: $\sqrt{(\sigma_{\varpi}\partial \chi^2/ \partial \varpi)^2 + (\sigma_{\mu\alpha}\partial \chi^2/ \partial \mu_{\alpha})^2 + \dots}$. The 6617 sources with a suspected IAD corruption are colored red, otherwise sources are blue. Upper panel: The results of the refit without \codename's ad-hoc correction. Lower panel: All five-parameter sources, after our ad-hoc correction. All of the former outliers that we applied our algorithm to (all but 38) are now well-behaved and a fit of their IAD yields the catalog solution (blue and red data points overlap).}
    \label{fig:refit_distributions}
\end{figure}

The top panel of Figure \ref{fig:refit_distributions} displays the quadrature sum of the chi-squared partials for all five-parameter sources, along with the difference between our refit proper motion and the catalog proper motions. The 6617 sources (colored in red) form a clear outlier cloud.

The culprit of the non-reproducibility of these 6617 sources appears to be data corruption in the IAD. Almost all of the discrepant sources have one or more lines of repeating AL errors at the end of their IAD file. The number of repeats, $n$, varies from source to source. For example, HIP 12241 has as the last four along-scan errors (in mas): 4.92, 4.10, 4.92, and 4.10. For HIP 42975 and 21000, the last ten along-scan errors are repeated in the same sequence. A second clue comes from the IAD header, which states the number of observations considered for the astrometric solution of a given source. For the problematic sources this number is $n$ less than the number of residual records available in the IAD file. This discrepancy persists even after removing transits marked as rejected by negative AL errors. An example is HIP 651, where the IAD header states 164 observations considered, but the residual records contain 165 transits. The mismatch between observations considered and residuals available is 1 for 5919 of these 6617 sources, 419 have $n=2$.

Deleting the last $n$ rows does not resolve the issue. We suspect instead that an additional $n$ transits were supposed to be removed at some stage of the astrometric processing, but were accidentally only partially removed: we suggest that these $n$ transits were removed from the data structures holding columns 6 and 7 of the IAD, but accidentally kept in the data structures corresponding to the first five columns. The repeating values of AL errors we observe could then be explained by access beyond the memory allocated for columns 6 and 7. Following this assumption, we developed an ad-hoc correction algorithm. This algorithm is described in detail in Appendix A.

The lower panel of Figure \ref{fig:refit_distributions} shows the new distribution of proper motion discrepancies and $\chi^2$ values after applying our ad-hoc correction algorithm to the IAD. The entire outlier cloud is removed, except for 38 sources where the ad-hoc correction is combinatorially infeasible. Refitting the modified IAD always recovers the catalog parameters. This strongly supports that our algorithm is correct and that the 6617 outliers that we discovered were caused by data corruption in their IAD.

We do not apply our ad-hoc correction to 38 sources because $n$ is so great for these sources that our ad-hoc correction requires an inordinately large computation time. GJ 660 (HIP 84123) is among these 38, which has $n = 21$ (which would take roughly 200 days on a single CPU core to correct). A list of the 38 outliers that we do not correct are listed in \texttt{htof/data/hip2\_Javatool\_flagged.txt}. \codename applies the ad-hoc correction automatically to any IAD from the Java tool, except these 38 sources. 

This fix is important for many uses of the IAD. Many \hipparcos sources host planetary or brown dwarf companions, and the masses of these companions can be measured by fitting the proper motion anomalies derived from five-parameter fits to the IAD (e.g., \citealp{Brandt_Dupuy_Bowler_2018, 2021arXiv210413941V}). Alternatively, many authors have fit the positions from the \hipparcos 2007 IAD directly in orbital analyses (e.g., \citealp{Sahlmann+Segransan+Queloz_2011, 2011AIPC.1331..102Q, Snellen+Brown_2018, 2019MNRAS_Fabo_Feng_etal_eps_indiA, Nielson+DeRosa+etal+betapicc2019, AMLagrange2020betapicc_direct_detection}). 
Future studies may include some of the 6617 sources. Without a correction of the data, the recovered proper motions can be systematically offset by a large fraction of the formal error. Such an offset would bias the inferred orbital parameters and companion masses. This risk can be mitigated by using \codename with the built-in ad-hoc correction.

The data corruption likewise has affected nearly the same subset of sources of the IAD on the \Hipparcos 2007 DVD. In principle, the same ad-hoc correction is possible. However, in practice it requires substantially more computation time, because rejected transits (those marked with negative uncertainties in the Java tool) are not clearly indicated in the DVD IAD. Even the information about the number of rejected observations in the IAD header only gives the integer part of the percentage of rejected observations. This means that it is not clearly indicated if any source with less than 100 observations has zero or one rejected observation. We recommend that any work using the IAD from the \Hipparcos 2007 DVD should exclude sources where the chi-squared partials of the data do not sum to zero. The DVD sources that \codename failed to refit are given in \texttt{htof/data/hip2\_DVD\_flagged.txt}. We defined a refit failure as one with proper motion refit discrepancies exceeding 0.03 mas/yr. These stars also have significantly nonzero sums of $\chi^2$ partials. We encourage the reader to consider using the IAD corresponding to the Java tool, where \codename has automatic correction built-in and so these data are ready for a variety of use cases like orbit fitting.

\section{Use Case I: Orbit Fitting}\label{sec:hipparcos_only_application}

\codename's primary use case to-date is in orbit fits that incorporate \gaia and \hipparcos absolute astrometry. \codename is implemented in \orbitcodename \citep{TimOrbitFitTemporary}, a Markov-Chain-Monte-Carlo (MCMC) code that employs the parallel-tempering sampler \texttt{ptemcee} \citep{Foreman-Mackey+Hogg+Lang+etal_2013,Vousden+Farr+Mandel_2016}. \codename is used to forward model the absolute astrometry within \orbitcodename. \codename and \orbitcodename were used, for example, to infer a precise mass for Gl~229~B \citep{brandt_gliese_229b_mass_htof, 2021_six_masses_GBrandt}; produce masses and orbital parameters for two white dwarf companions \citep{2021AJ....161..106B}; yield a dynamical mass estimate for a newly discovered brown dwarf companion to HD~33632 \citep{2020ApJ_Currie_Thayne_HD33632}; infer the mass of a substellar companion to HD~47127 \citep{Bowler2021accepted}; produce refined masses and orbital parameters for $\beta$~Pic~b and c \citep{Brandt2020betapicbc}; \addition{derive the first dynamical mass for HR~8799~e \citep{2021_Brandtetal_HR8799}}; derive precise masses for six directly imaged brown dwarfs \citep{2021_six_masses_GBrandt}; and solve for masses (breaking the inclination degeneracy) of nine massive companions that previously had only RV detections \citep{2021_Li_Yiting_nine_masses_RV_planets}. 

\codename uses the \gaia scanning law, \hipparcos 1997 and 2007 IAD to forward model perturbations to the absolute astrometry due to companions. \codename computes three proper motion perturbations per MCMC step: one for \gaia, one for \hipparcos, and finally one to the HGCA long-baseline proper motion. The unperturbed proper motions are cross-calibrated and have already been placed in a common reference frame \citep{brandt_cross_cal_gaia_2018, 2021HGCA_edr3_Brandt}. \codename's forward modelling of the absolute astrometry is particularly relevant for sources where the precision on the mass of the companion is high. For Gl~229~B, with a mass measured to $\lesssim$1\%, disabling \codename's exact forward modelling causes the inferred mass of the companion to shift by an amount comparable to its formal uncertainty \citep{2021_six_masses_GBrandt}.

\section{Use Case II: Astrometric mission forecasting}\label{sec:mission_forecasting}

Precision astrometric missions can detect and characterize companions around stars by observing the reflex motion of the star in the plane of the sky due to its companion. \gaia expects to detect tens of thousands of sub-stellar companions over its nominal 5-year mission \citep{Perryman+Hartman+Bakos+etal_2014}. However, many additional missions have been proposed (in part) for exoplanet discovery and characterization. For example SIM and SIM Lite (which aimed at roughly $\mu$as precision; \citealp{2008PASPSIM_PlanetQuest, 2009_SimLITE}), JASMINE (formerly named Small-JASMINE;\footnote{\url{http://jasmine.nao.ac.jp/index-en.html} and \\ \url{http://www.scholarpedia.org/article/JASMINE}} \citealt{gouda_2020}), TOLIMAN \citep{2021inas.book..167D_TOLIMAN}, Theia \citep{2017arXiv170701348T, 2019arXiv191008028M}, and \ngrst \citep{wfirst_spergel_2015, Melchor+Spergel+Lanz_2018}. Here we show how \codename can calculate new planets that would be discoverable in the $\beta$~Pic system if one combined \hipparcos, \gaia and \ngrst measurements. The example discussed in this section is contained in the Jupyter examples notebook that accompanies \codename. \codename's mission forecasting capability is applicable to any combination of astrometric missions for any star. 

\subsection{Data and assumptions}

For \ngrst, we use the 10\,$\mu$as precision estimated by \cite{Sanderson+Bellini+etal_WFIRSTprecision} for spatial scanning (single scan). \cite{Melchor+Spergel+Lanz_2018} estimate a $\sim$10\,$\mu$as precision for bright stars from centering on diffraction spikes. The end-of-mission performance of \Gaia on bright stars is expected to be a factor of a few better than it is in EDR3. The exact end-of-mission single-scan precision for \Gaia is unavailable but we estimate it as follows. The \Gaia collaboration predicts 10\,$\mu$as parallax precision by \gaia DR4 for bright stars ($3 < V < 12$).\footnote{\url{https://www.cosmos.esa.int/web/gaia/science-performance}} The error on the sky-averaged position is 0.75$\times$ the parallax precision. Each source will be observed $\sim$100 times. A rough estimate of the single-exposure precision is then $\sim 10 \cdot 0.75 \sqrt{100}\,\mu$as~$ = 75\,\mu$as. We inflate this slightly, to be conservative, to $120\,\mu$as.

We now consider which planets are detectable given three different combinations of data on $\beta$~Pic: \Gaia alone (which we label as \gaia), \Gaia combined with \Hipparcos ({\it Gaia}\,+\,HIP), and \Gaia combined with a mock \ngrst campaign ({\it Gaia}\,+\,{\sl NGRST}). The mock \ngrst campaign of $\beta$~Pic consists of six randomly spaced \ngrst spatial-scanning observations between 2025 and 2028. We assume a diagonal covariance matrix for each \ngrst observation with 10\,$\mu$as errors in right-ascension and declination. The mock \Gaia DR4 data are the 91 \Gaia GOST observations planned for $\beta$~Pic between 2014 and 2020. We assume equal along-scan errors of 120$\mu$as and zero covariance between measurements. The \Hipparcos data we use are the actual \Hipparcos observations of $\beta$~Pic from the 1997 reduction IAD. 

We compute the $\Delta \chi^2$ of a linear astrometric fit to the perturbed motion of $\beta$~Pic for a swath of planetary masses and semi-major axes. For any given mass and semi-major axis, we assume a circular edge-on orbit ($i = 90\degree$) around $\beta$~Pic~A. We assume that effects like apparent acceleration from projected linear motion and timing fluctuations from light travel time have been accounted for. Finally, we take the average $\Delta \chi^2$ across a uniform distribution of orbital phases. We define a body as detectable if it causes an astrometric perturbation on its host star with $\Delta \chi^2 > 30$ where $\Delta \chi^2 = 0$ implies linear motion perfectly described by five astrometric parameters. $\Delta \chi^2 > 30$ is the `marginal' detection criterion from \cite{Perryman+Hartman+Bakos+etal_2014}. $\Delta \chi^2 > 100$ is a strong detection where orbital parameters can typically be constrained to better than 10\% \citep{Perryman+Hartman+Bakos+etal_2014}.

\subsection{Results and discussion}

\begin{figure}[t!]
    \centering
    \includegraphics[width=\linewidth]{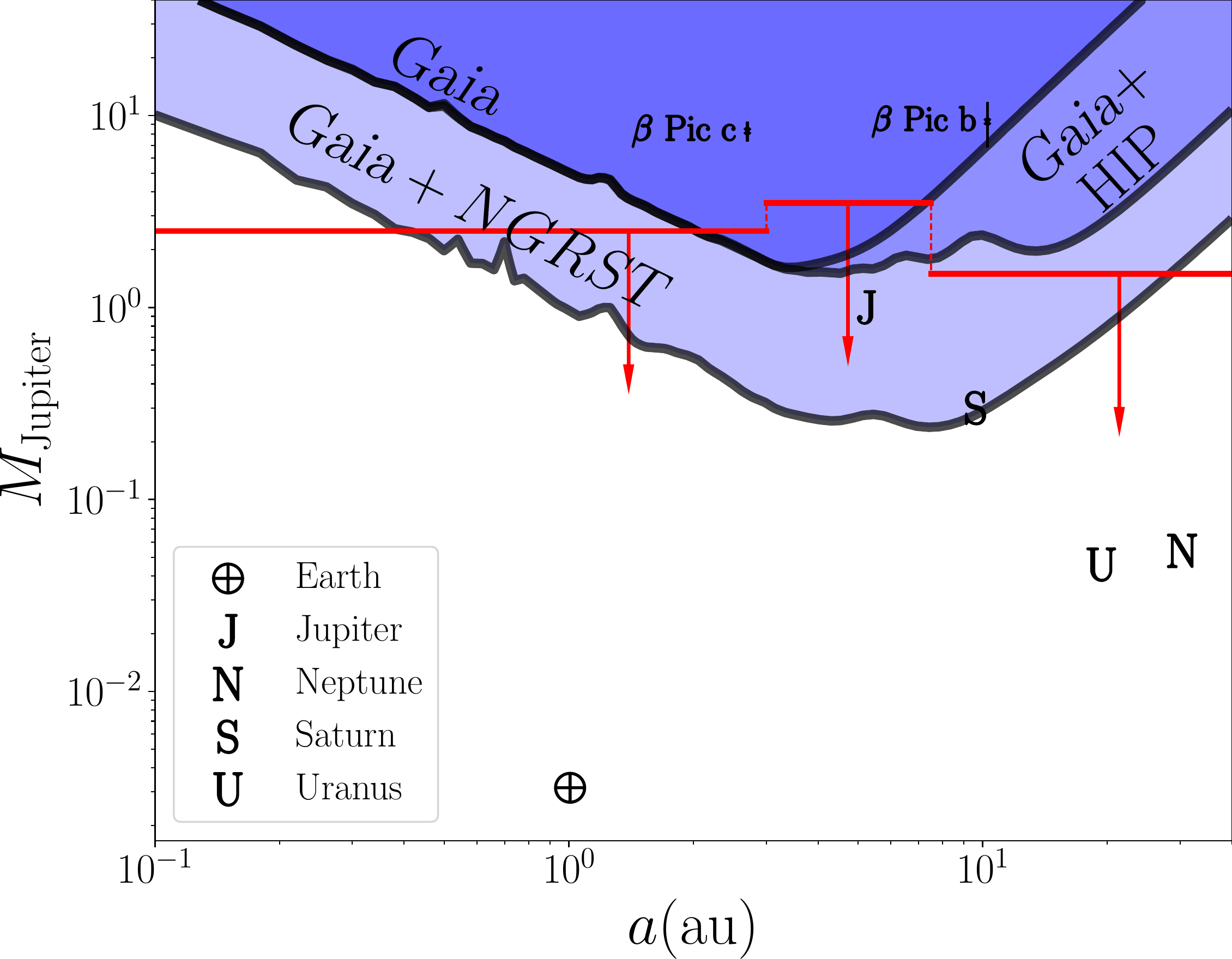}
    \caption{Mass and semimajor axis space for planets orbiting the star $\beta$~Pic~A in circular, edge-on orbits. Planets $\beta$~Pic~b and $\beta$~Pic~c have been placed according to their semi-major axis' and masses \citep{Brandt2020betapicbc}. Solar system planets are shown for comparison. Blue regions indicate parameter space where a planet would be detectable with $\Delta \chi^2 > 30$ --- corresponding to the \cite{Perryman+Hartman+Bakos+etal_2014} detection criterion. Planets in any blue region (above the {\sl Gaia}\,+\,{\sl NGRST} contour) would be detectable by combining \Gaia DR4 with six \ngrst observations. Planets in the dark-blue region above the {\sl Gaia} contour would be detectable by \Gaia alone, and planets in either of the darkest two regions ({\sl Gaia} or {\sl Gaia}\,+\,HIP) could be discovered through a merging of the \Gaia and \Hipparcos IAD. The regions above the red lines correspond to additional planets ruled out by \cite{AMLagrange2020betapicc_direct_detection}. The code to reproduce the {\sl Gaia}\,+\,{\sl NGRST} contour is in the Jupyter examples notebook that accompanies \codename.}
    \label{fig:gaia_wfirst_betapic}
\end{figure}

Figure \ref{fig:gaia_wfirst_betapic} shows the parameter space of hypothetical planets around $\beta$~Pic~A that are detectable. The known planets $\beta$~Pic~b and $\beta$~Pic~c are shown. We show contours of $\Delta \chi^2 > 30$ in mass and semi-major axis space for $\beta$~Pic for all three cases: \gaia, {\it Gaia}+{\sl NGRST}, and {\it Gaia}+HIP. \Hipparcos will be important for potential short-period planets in $\beta$~Pic only if the \Gaia end-of-mission parallax precision is substantially worse than 16\,$\mu$as. In the {\it Gaia}+{\sl NGRST} case (all blue regions), adding \Hipparcos astrometry negligibly enlarges the detectable region because of the already long time baseline between \Gaia and \ngrst and the high precision of \ngrst. $\beta$~Pic~c is detectable with confidence with Gaia DR4 alone ($\Delta \chi^2$ exceeds 400) and with extreme confidence ($\Delta \chi^2$ exceeds 1000) if combined with six \ngrst observations. An unseen Saturn mass companion could be detected by combining \Gaia and \ngrst.

\cite{AMLagrange2020betapicc_direct_detection} ruled out the presence of additional planets in the $\beta$~Pic system more massive than $\sim$2.5\,$\Mjup$ closer than 3~au; 3.5\,$\Mjup$ for planets whose semi major axis' are between 3~au and 7.5~au; and $\sim$1-2$\Mjup$ for planets lying beyond 7.5 au. Figure \ref{fig:gaia_wfirst_betapic} shows these limits as red lines. We find that a \Gaia DR4 and \Hipparcos merger could improve the limit between 3~au and 7.5~au to exclude additional planets (in edge-on orbits) more massive than 1.5\,$\Mjup$. Adding six \ngrst observations could exclude edge-on planets more massive than 0.2\,$\Mjup$.

\section{Summary and Conclusions}\label{sec:conclusions}
We present \codename, an open-source tool for using the \Hipparcos and \Gaia astrometry in a variety of applications. \codename provides routines to parse and fit the intermediate astrometric data from both \Hipparcos reductions and the \gaia scanning law. It is prepared for future \Gaia data releases which will include the intermediate astrometric data and higher-order fits. 

We suggest that 6617 sources within \hipparcos 2007 have corrupted IAD, and present \codename's ad-hoc correction to this corruption (for the Java tool version of the IAD). \codename can be used to forecast future astrometric missions. We find that \Gaia DR4 merged with \hipparcos could improve the \cite{AMLagrange2020betapicc_direct_detection} $\beta$~Pic companion limits by a factor of 2, excluding additional planets more massive than $\sim$1.5$\Mjup$ with semi-major axis' between 3~au and 7.5~au. Jupiter and Saturn mass companions would be astrometrically detectable around $\beta$~Pic if one combined \Gaia DR4 with six \ngrst observations.

\codename is available at \url{https://github.com/gmbrandt/htof} and installable with \texttt{pip install htof}. The \texttt{examples.ipynb} Jupyter notebook provides examples of using \codename. We welcome feedback and contributions.

\section*{acknowledgements}
We thank the referee for helpful feedback on this work. We thank Anthony G.~A.~Brown for helpful discussions about \Gaia and \Hipparcos, and for the sky path module in \codename which we obtained, with his permission, from his package \texttt{astrometric-sky-path} \footnote{\url{https://github.com/agabrown/astrometric-sky-path}}.
We thank Floor van Leeuwen, for supporting us with access to the IAD from the Java tool, for helping us interpret the Java tool IAD and some of their detailed properties, and for helpful discussions of the manuscript. G.~M.~B. thanks Lachlan T. Lancaster for helpful discussions. 

D.~M. gratefully acknowledges support from the European Space Agency (ESA) Research Fellow programme. G.~M.~B. is supported by the National Science Foundation (NSF) Graduate Research Fellowship under grant no. 1650114.

This work made use of the \texttt{REBOUND} code which is freely available at http://github.com/hannorein/rebound.
This work has made use of data from the European Space Agency (ESA) mission \gaia (\url{https://www.cosmos.esa.int/Gaia}), processed by the \gaia Data Processing and Analysis Consortium (DPAC, \url{https://www.cosmos.esa.int/web/Gaia/dpac/consortium}). Funding for the DPAC has been provided by national institutions, in particular the institutions participating in the \gaia Multilateral Agreement.

\software{astropy \citep{astropy:2013, astropy:2018}, 
          scipy \citep{2020SciPy-NMeth},
          numpy \citep{numpy1, numpy2},
          pandas \citep{mckinney-proc-scipy-2010, reback2020pandas},
          Jupyter, 
          }
          
\appendix

\section{The Ad-hoc Correction Algorithm}
Here we detail the ad-hoc correction algorithm for the \hipparcos 2007 Java tool IAD. The Jupyter notebook with the source code contains examples that demonstrate the ad-hoc correction.
\begin{compactenum}
    \item We separate the IAD into two blocks: the first five columns (orbit number, epoch, parallax factor, sine and cosine of the scan angle) into the left-hand side block L; and the last two columns (residuals and AL errors) into the right-hand side block R.
    \item We assume that the last $n$ residuals and (typically duplicate) along-scan errors are corrupted and therefore remove the last $n$ rows from block R. This results in $n$ too many rows in block L. 
    \item We iterate through all possible combinations of removing $n$ rows from block L.\footnote{We speed up the algorithm for the 180 sources with $n \geq 5$. These have a large number of combinatoric possibilities. Often, one orbit contains multiple observations. These often almost have identical epochs and scan angles. For all practical purposes, we find that it does not matter which one of the observations within one orbit we remove, as long as we remove the correct number of observations from that orbit. }
    For each iteration:
    \begin{compactenum}
        \item Blocks L and R now have the same number of rows again. We recombine blocks L and R, yielding a modified IAD.
        \item We calculate the quadrature sum of the chi-squared partials for this modified IAD.
    \end{compactenum}
    \item We accept the modified IAD that gives the smallest quadrature sum of the chi-squared partials. This sum is typically less than 0.01. 
\end{compactenum}

\bibliography{refs}{}

\begin{thebibliography}{}
\expandafter\ifx\csname natexlab\endcsname\relax\def\natexlab#1{#1}\fi
\providecommand{\url}[1]{\href{#1}{#1}}
\providecommand{\dodoi}[1]{doi:~\href{http://doi.org/#1}{\nolinkurl{#1}}}
\providecommand{\doeprint}[1]{\href{http://ascl.net/#1}{\nolinkurl{http://ascl.net/#1}}}
\providecommand{\doarXiv}[1]{\href{https://arxiv.org/abs/#1}{\nolinkurl{https://arxiv.org/abs/#1}}}

\bibitem[{{Astropy Collaboration} {et~al.}(2013){Astropy Collaboration},
  {Robitaille}, {Tollerud}, {Greenfield}, {Droettboom}, {Bray}, {Aldcroft},
  {Davis}, {Ginsburg}, {Price-Whelan}, {Kerzendorf}, {Conley}, {Crighton},
  {Barbary}, {Muna}, {Ferguson}, {Grollier}, {Parikh}, {Nair}, {Unther},
  {Deil}, {Woillez}, {Conseil}, {Kramer}, {Turner}, {Singer}, {Fox}, {Weaver},
  {Zabalza}, {Edwards}, {Azalee Bostroem}, {Burke}, {Casey}, {Crawford},
  {Dencheva}, {Ely}, {Jenness}, {Labrie}, {Lim}, {Pierfederici}, {Pontzen},
  {Ptak}, {Refsdal}, {Servillat}, \& {Streicher}}]{astropy:2013}
{Astropy Collaboration}, {Robitaille}, T.~P., {Tollerud}, E.~J., {et~al.} 2013,
  \aap, 558, A33, \dodoi{10.1051/0004-6361/201322068}

\bibitem[{{Bowler} {et~al.}(2021{\natexlab{a}}){Bowler}, {Cochran}, {Endl},
  {Franson}, {Brandt}, {Dupuy}, {MacQueen}, {Kratter}, {Mawet}, \&
  {Ruane}}]{2021AJ....161..106B}
{Bowler}, B.~P., {Cochran}, W.~D., {Endl}, M., {et~al.} 2021{\natexlab{a}},
  \aj, 161, 106, \dodoi{10.3847/1538-3881/abd243}

\bibitem[{{Bowler} {et~al.}(2021{\natexlab{b}}){Bowler}, {Endl}, {Cochran},
  {MacQueen}, {Crepp}, {Doppmann}, {Dulz}, {Brandt}, {Brandt}, {Li}, {Dupuy},
  {Franson}, {Kratter}, {Morley}, \& {Zhou}}]{Bowler2021accepted}
{Bowler}, B.~P., {Endl}, M., {Cochran}, W.~D., {et~al.} 2021{\natexlab{b}},
  arXiv e-prints, arXiv:2105.01255.
\newblock \doarXiv{2105.01255}

\bibitem[{{Brandt} {et~al.}(2021{\natexlab{a}}){Brandt}, {Brandt}, {Dupuy},
  {Li}, \& {Michalik}}]{Brandt2020betapicbc}
{Brandt}, G.~M., {Brandt}, T.~D., {Dupuy}, T.~J., {Li}, Y., \& {Michalik}, D.
  2021{\natexlab{a}}, \aj, 161, 179, \dodoi{10.3847/1538-3881/abdc2e}

\bibitem[{{Brandt} {et~al.}(2021{\natexlab{b}}){Brandt}, {Brandt}, {Dupuy},
  {Michalik}, \& {Marleau}}]{2021_Brandtetal_HR8799}
{Brandt}, G.~M., {Brandt}, T.~D., {Dupuy}, T.~J., {Michalik}, D., \& {Marleau},
  G.-D. 2021{\natexlab{b}}, \apjl, 915, L16, \dodoi{10.3847/2041-8213/ac0540}

\bibitem[{Brandt \& Michalik(2021)}]{htof_zenodo}
Brandt, G.~M., \& Michalik, D. 2021, HTOF, 1.0.0,  Zenodo,
  \dodoi{10.5281/zenodo.5495866}

\bibitem[{{Brandt} {et~al.}(2021{\natexlab{c}}){Brandt}, {Dupuy}, {Li}, {Chen},
  {Brandt}, {Wong}, {Currie}, {Bowler}, {Liu}, {Best}, \&
  {Phillips}}]{2021_six_masses_GBrandt}
{Brandt}, G.~M., {Dupuy}, T.~J., {Li}, Y., {et~al.} 2021{\natexlab{c}}, arXiv
  e-prints, arXiv:2109.07525.
\newblock \doarXiv{2109.07525}

\bibitem[{{Brandt}(2018)}]{brandt_cross_cal_gaia_2018}
{Brandt}, T.~D. 2018, The Astrophysical Journal Supplement Series, 239, 31,
  \dodoi{10.3847/1538-4365/aaec06}

\bibitem[{{Brandt}(2021)}]{2021HGCA_edr3_Brandt}
---. 2021, arXiv e-prints, arXiv:2105.11662.
\newblock \doarXiv{2105.11662}

\bibitem[{{Brandt} {et~al.}(2018){Brandt}, {Dupuy}, \&
  {Bowler}}]{Brandt_Dupuy_Bowler_2018}
{Brandt}, T.~D., {Dupuy}, T.~J., \& {Bowler}, B.~P. 2018, arXiv e-prints,
  arXiv:1811.07285.
\newblock \doarXiv{1811.07285}

\bibitem[{Brandt {et~al.}(2019)Brandt, Dupuy, Bowler, Gagliuffi, Faherty,
  Brandt, \& Michalik}]{brandt_gliese_229b_mass_htof}
Brandt, T.~D., Dupuy, T.~J., Bowler, B.~P., {et~al.} 2019, A Dynamical Mass of
  $70 \pm 5$ Jupiter Masses for Gliese 229B, the First Imaged T Dwarf.
\newblock \doarXiv{1910.01652}

\bibitem[{{Brandt} {et~al.}(2021{\natexlab{d}}){Brandt}, {Dupuy}, {Li},
  {Brandt}, {Zeng}, {Michalik}, {Bardalez Gagliuffi}, \&
  {Raposo-Pulido}}]{TimOrbitFitTemporary}
{Brandt}, T.~D., {Dupuy}, T.~J., {Li}, Y., {et~al.} 2021{\natexlab{d}}, arXiv
  e-prints, arXiv:2105.11671.
\newblock \doarXiv{2105.11671}

\bibitem[{{Calissendorff} \& {Janson}(2018)}]{Calissendorff+Janson_2018}
{Calissendorff}, P., \& {Janson}, M. 2018, ArXiv e-prints.
\newblock \doarXiv{1806.07899}

\bibitem[{{Currie} {et~al.}(2020){Currie}, {Brandt}, {Kuzuhara}, {Chilcote},
  {Guyon}, {Marois}, {Groff}, {Lozi}, {Vievard}, {Sahoo}, {Deo}, {Jovanovic},
  {Martinache}, {Wagner}, {Dupuy}, {Wahl}, {Letawsky}, {Li}, {Zeng}, {Brandt},
  {Michalik}, {Grady}, {Janson}, {Knapp}, {Kwon}, {Lawson}, {McElwain},
  {Uyama}, {Wisniewski}, \& {Tamura}}]{2020ApJ_Currie_Thayne_HD33632}
{Currie}, T., {Brandt}, T.~D., {Kuzuhara}, M., {et~al.} 2020, \apjl, 904, L25,
  \dodoi{10.3847/2041-8213/abc631}

\bibitem[{{de Bruijne, J. H. J.} \& {Eilers, A.-C.}(2012)}]{htpm_JHJ_2012AA}
{de Bruijne, J. H. J.}, \& {Eilers, A.-C.} 2012, A\&A, 546, A61,
  \dodoi{10.1051/0004-6361/201219219}

\bibitem[{{De Rosa} {et~al.}(2020){De Rosa}, {Dawson}, \&
  {Nielsen}}]{2020A&A...640A..73D}
{De Rosa}, R.~J., {Dawson}, R., \& {Nielsen}, E.~L. 2020, \aap, 640, A73,
  \dodoi{10.1051/0004-6361/202038496}

\bibitem[{{Delli Veneri} {et~al.}(2021){Delli Veneri}, {Desdoigts}, {Schmitz},
  {Krone-Martins}, {Ishida}, {Tuthill}, {de Souza}, {Scalzo}, {Brescia},
  {Longo}, \& {Picariello}}]{2021inas.book..167D_TOLIMAN}
{Delli Veneri}, M., {Desdoigts}, L., {Schmitz}, M.~A., {et~al.} 2021, {Periodic
  Astrometric Signal Recovery Through Convolutional Autoencoders}, Vol.~39,
  167--195, \dodoi{10.1007/978-3-030-65867-0\_8}

\bibitem[{{Dupuy} {et~al.}(2019){Dupuy}, {Brandt}, {Kratter}, \&
  {Bowler}}]{Dupuy+Brandt+Kratter+etal_2019}
{Dupuy}, T.~J., {Brandt}, T.~D., {Kratter}, K.~M., \& {Bowler}, B.~P. 2019,
  \apjl, 871, L4, \dodoi{10.3847/2041-8213/aafb31}

\bibitem[{ESA(accessed February 4, 2020)}]{gaia_scanning_law_info}
ESA. accessed February 4, 2020,
  \url{https://www.cosmos.esa.int/web/gaia/scanning-law}

\bibitem[{{Feng} {et~al.}(2019){Feng}, {Anglada-Escud{\'e}}, {Tuomi}, {Jones},
  {Chanam{\'e}}, {Butler}, \& {Janson}}]{2019MNRAS_Fabo_Feng_etal_eps_indiA}
{Feng}, F., {Anglada-Escud{\'e}}, G., {Tuomi}, M., {et~al.} 2019, \mnras, 490,
  5002, \dodoi{10.1093/mnras/stz2912}

\bibitem[{Fern\'andez-Hern\'andez \& Joliet(2019)}]{gaia_gost_user_manual}
Fern\'andez-Hern\'andez, J., \& Joliet, E. 2019, {GOST} Software User Manual,
  Gaia DPAC

\bibitem[{{Fontanive} {et~al.}(2019){Fontanive}, {Mu{\v{z}}i{\'c}}, {},
  {Bonavita}, \& {Biller}}]{2019MNRAS.490.1120F}
{Fontanive}, C., {Mu{\v{z}}i{\'c}}, {}, K., {Bonavita}, M., \& {Biller}, B.
  2019, \mnras, 490, 1120, \dodoi{10.1093/mnras/stz2587}

\bibitem[{{Foreman-Mackey} {et~al.}(2013){Foreman-Mackey}, {Hogg}, {Lang}, \&
  {Goodman}}]{Foreman-Mackey+Hogg+Lang+etal_2013}
{Foreman-Mackey}, D., {Hogg}, D.~W., {Lang}, D., \& {Goodman}, J. 2013, \pasp,
  125, 306, \dodoi{10.1086/670067}

\bibitem[{{Gaia Collaboration} {et~al.}(2020){Gaia Collaboration}, {Brown},
  {Vallenari}, {Prusti}, {de Bruijne}, {Babusiaux}, \&
  {Biermann}}]{2020GaiaEDR3_catalog_summary}
{Gaia Collaboration}, {Brown}, A.~G.~A., {Vallenari}, A., {et~al.} 2020, arXiv
  e-prints, arXiv:2012.01533.
\newblock \doarXiv{2012.01533}

\bibitem[{{Gaia Collaboration} {et~al.}(2016){Gaia Collaboration}, {Prusti},
  {de Bruijne}, {Brown}, {Vallenari}, {Babusiaux}, {Bailer-Jones}, {Bastian},
  {Biermann}, {Evans}, {Eyer}, {Jansen}, {Jordi}, {Klioner}, {Lammers},
  {Lindegren}, {Luri}, {Mignard}, {Milligan}, {Panem}, {Poinsignon},
  {Pourbaix}, {Randich}, {Sarri}, {Sartoretti}, {Siddiqui}, {Soubiran},
  {Valette}, {van Leeuwen}, {Walton}, {Aerts}, {Arenou}, {Cropper}, {Drimmel},
  {H{\o}g}, {Katz}, {Lattanzi}, {O'Mullane}, {Grebel}, {Holland}, {Huc},
  {Passot}, {Bramante}, {Cacciari}, {Casta{\~n}eda}, {Chaoul}, {Cheek}, {De
  Angeli}, {Fabricius}, {Guerra}, {Hern{\'a}ndez}, {Jean-Antoine-Piccolo},
  {Masana}, {Messineo}, {Mowlavi}, {Nienartowicz}, {Ord{\'o}{\~n}ez- Blanco},
  {Panuzzo}, {Portell}, {Richards}, {Riello}, {Seabroke}, {Tanga},
  {Th{\'e}venin}, {Torra}, {Els}, {Gracia- Abril}, {Comoretto},
  {Garcia-Reinaldos}, {Lock}, {Mercier}, {Altmann}, {Andrae}, {Astraatmadja},
  {Bellas-Velidis}, {Benson}, {Berthier}, {Blomme}, {Busso}, {Carry},
  {Cellino}, {Clementini}, {Cowell}, {Creevey}, {Cuypers}, {Davidson}, {De
  Ridder}, {de Torres}, {Delchambre}, {Dell'Oro}, {Ducourant}, {Fr{\'e}mat},
  {Garc{\'\i}a-Torres}, {Gosset}, {Halbwachs}, {Hambly}, {Harrison}, {Hauser},
  {Hestroffer}, {Hodgkin}, {Huckle}, {Hutton}, {Jasniewicz}, {Jordan},
  {Kontizas}, {Korn}, {Lanzafame}, {Manteiga}, {Moitinho}, {Muinonen},
  {Osinde}, {Pancino}, {Pauwels}, {Petit}, {Recio-Blanco}, {Robin}, {Sarro},
  {Siopis}, {Smith}, {Smith}, {Sozzetti}, {Thuillot}, {van Reeven}, {Viala},
  {Abbas}, {Abreu Aramburu}, {Accart}, {Aguado}, {Allan}, {Allasia},
  {Altavilla}, {{\'A}lvarez}, {Alves}, {Anderson}, {Andrei}, {Anglada Varela},
  {Antiche}, {Antoja}, {Ant{\'o}n}, {Arcay}, {Atzei}, {Ayache}, {Bach},
  {Baker}, {Balaguer-N{\'u}{\~n}ez}, {Barache}, {Barata}, {Barbier}, {Barblan},
  {Baroni}, {Barrado y Navascu{\'e}s}, {Barros}, {Barstow}, {Becciani},
  {Bellazzini}, {Bellei}, {Bello Garc{\'\i}a}, {Belokurov}, {Bendjoya},
  {Berihuete}, {Bianchi}, {Bienaym{\'e}}, {Billebaud}, {Blagorodnova},
  {Blanco-Cuaresma}, {Boch}, {Bombrun}, {Borrachero}, {Bouquillon}, {Bourda},
  {Bouy}, {Bragaglia}, {Breddels}, {Brouillet}, {Br{\"u}semeister},
  {Bucciarelli}, {Budnik}, {Burgess}, {Burgon}, {Burlacu}, {Busonero}, {Buzzi},
  {Caffau}, {Cambras}, {Campbell}, {Cancelliere}, {Cantat-Gaudin}, {Carlucci},
  {Carrasco}, {Castellani}, {Charlot}, {Charnas}, {Charvet}, {Chassat},
  {Chiavassa}, {Clotet}, {Cocozza}, {Collins}, {Collins}, {Costigan}, {Crifo},
  {Cross}, {Crosta}, {Crowley}, {Dafonte}, {Damerdji}, {Dapergolas}, {David},
  {David}, {De Cat}, {de Felice}, {de Laverny}, {De Luise}, {De March}, {de
  Martino}, {de Souza}, {Debosscher}, {del Pozo}, {Delbo}, {Delgado},
  {Delgado}, {di Marco}, {Di Matteo}, {Diakite}, {Distefano}, {Dolding}, {Dos
  Anjos}, {Drazinos}, {Dur{\'a}n}, {Dzigan}, {Ecale}, {Edvardsson}, {Enke},
  {Erdmann}, {Escolar}, {Espina}, {Evans}, {Eynard Bontemps}, {Fabre},
  {Fabrizio}, {Faigler}, {Falc{\~a}o}, {Farr{\`a}s Casas}, {Faye}, {Federici},
  {Fedorets}, {Fern{\'a}ndez-Hern{\'a}ndez}, {Fernique}, {Fienga}, {Figueras},
  {Filippi}, {Findeisen}, {Fonti}, {Fouesneau}, {Fraile}, {Fraser}, {Fuchs},
  {Furnell}, {Gai}, {Galleti}, {Galluccio}, {Garabato}, {Garc{\'\i}a-Sedano},
  {Gar{\'e}}, {Garofalo}, {Garralda}, {Gavras}, {Gerssen}, {Geyer}, {Gilmore},
  {Girona}, {Giuffrida}, {Gomes}, {Gonz{\'a}lez-Marcos},
  {Gonz{\'a}lez-N{\'u}{\~n}ez}, {Gonz{\'a}lez-Vidal}, {Granvik}, {Guerrier},
  {Guillout}, {Guiraud}, {G{\'u}rpide}, {Guti{\'e}rrez-S{\'a}nchez}, {Guy},
  {Haigron}, {Hatzidimitriou}, {Haywood}, {Heiter}, {Helmi}, {Hobbs},
  {Hofmann}, {Holl}, {Holland}, {Hunt}, {Hypki}, {Icardi}, {Irwin}, {Jevardat
  de Fombelle}, {Jofr{\'e}}, {Jonker}, {Jorissen}, {Julbe}, {Karampelas},
  {Kochoska}, {Kohley}, {Kolenberg}, {Kontizas}, {Koposov}, {Kordopatis},
  {Koubsky}, {Kowalczyk}, {Krone-Martins}, {Kudryashova}, {Kull}, {Bachchan},
  {Lacoste-Seris}, {Lanza}, {Lavigne}, {Le Poncin-Lafitte}, {Lebreton},
  {Lebzelter}, {Leccia}, {Leclerc}, {Lecoeur-Taibi}, {Lemaitre}, {Lenhardt},
  {Leroux}, {Liao}, {Licata}, {Lindstr{\o}m}, {Lister}, {Livanou}, {Lobel},
  {L{\"o}ffler}, {L{\'o}pez}, {Lopez-Lozano}, {Lorenz}, {Loureiro},
  {MacDonald}, {Magalh{\~a}es Fernandes}, {Managau}, {Mann}, {Mantelet},
  {Marchal}, {Marchant}, {Marconi}, {Marie}, {Marinoni}, {Marrese},
  {Marschalk{\'o}}, {Marshall}, {Mart{\'\i}n-Fleitas}, {Martino}, {Mary},
  {Matijevi{\v{c}}}, {Mazeh}, {McMillan}, {Messina}, {Mestre}, {Michalik},
  {Millar}, {Miranda}, {Molina}, {Molinaro}, {Molinaro}, {Moln{\'a}r},
  {Moniez}, {Montegriffo}, {Monteiro}, {Mor}, {Mora}, {Morbidelli}, {Morel},
  {Morgenthaler}, {Morley}, {Morris}, {Mulone}, {Muraveva}, {Musella},
  {Narbonne}, {Nelemans}, {Nicastro}, {Noval}, {Ord{\'e}novic},
  {Ordieres-Mer{\'e}}, {Osborne}, {Pagani}, {Pagano}, {Pailler}, {Palacin},
  {Palaversa}, {Parsons}, {Paulsen}, {Pecoraro}, {Pedrosa}, {Pentik{\"a}inen},
  {Pereira}, {Pichon}, {Piersimoni}, {Pineau}, {Plachy}, {Plum}, {Poujoulet},
  {Pr{\v{s}}a}, {Pulone}, {Ragaini}, {Rago}, {Rambaux}, {Ramos-Lerate},
  {Ranalli}, {Rauw}, {Read}, {Regibo}, {Renk}, {Reyl{\'e}}, {Ribeiro},
  {Rimoldini}, {Ripepi}, {Riva}, {Rixon}, {Roelens}, {Romero-G{\'o}mez},
  {Rowell}, {Royer}, {Rudolph}, {Ruiz-Dern}, {Sadowski}, {Sagrist{\`a}
  Sell{\'e}s}, {Sahlmann}, {Salgado}, {Salguero}, {Sarasso}, {Savietto},
  {Schnorhk}, {Schultheis}, {Sciacca}, {Segol}, {Segovia}, {Segransan},
  {Serpell}, {Shih}, {Smareglia}, {Smart}, {Smith}, {Solano}, {Solitro},
  {Sordo}, {Soria Nieto}, {Souchay}, {Spagna}, {Spoto}, {Stampa}, {Steele},
  {Steidelm{\"u}ller}, {Stephenson}, {Stoev}, {Suess}, {S{\"u}veges}, {Surdej},
  {Szabados}, {Szegedi-Elek}, {Tapiador}, {Taris}, {Tauran}, {Taylor},
  {Teixeira}, {Terrett}, {Tingley}, {Trager}, {Turon}, {Ulla}, {Utrilla},
  {Valentini}, {van Elteren}, {Van Hemelryck}, {van Leeuwen}, {Varadi},
  {Vecchiato}, {Veljanoski}, {Via}, {Vicente}, {Vogt}, {Voss}, {Votruba},
  {Voutsinas}, {Walmsley}, {Weiler}, {Weingrill}, {Werner}, {Wevers},
  {Whitehead}, {Wyrzykowski}, {Yoldas}, {{\v{Z}}erjal}, {Zucker}, {Zurbach},
  {Zwitter}, {Alecu}, {Allen}, {Allende Prieto}, {Amorim},
  {Anglada-Escud{\'e}}, {Arsenijevic}, {Azaz}, {Balm}, {Beck}, {Bernstein},
  {Bigot}, {Bijaoui}, {Blasco}, {Bonfigli}, {Bono}, {Boudreault}, {Bressan},
  {Brown}, {Brunet}, {Bunclark}, {Buonanno}, {Butkevich}, {Carret}, {Carrion},
  {Chemin}, {Ch{\'e}reau}, {Corcione}, {Darmigny}, {de Boer}, {de Teodoro}, {de
  Zeeuw}, {Delle Luche}, {Domingues}, {Dubath}, {Fodor}, {Fr{\'e}zouls},
  {Fries}, {Fustes}, {Fyfe}, {Gallardo}, {Gallegos}, {Gardiol}, {Gebran},
  {Gomboc}, {G{\'o}mez}, {Grux}, {Gueguen}, {Heyrovsky}, {Hoar}, {Iannicola},
  {Isasi Parache}, {Janotto}, {Joliet}, {Jonckheere}, {Keil}, {Kim},
  {Klagyivik}, {Klar}, {Knude}, {Kochukhov}, {Kolka}, {Kos}, {Kutka}, {Lainey},
  {LeBouquin}, {Liu}, {Loreggia}, {Makarov}, {Marseille}, {Martayan},
  {Martinez-Rubi}, {Massart}, {Meynadier}, {Mignot}, {Munari}, {Nguyen},
  {Nordlander}, {Ocvirk}, {O'Flaherty}, {Olias Sanz}, {Ortiz}, {Osorio},
  {Oszkiewicz}, {Ouzounis}, {Palmer}, {Park}, {Pasquato}, {Peltzer}, {Peralta},
  {P{\'e}turaud}, {Pieniluoma}, {Pigozzi}, {Poels}, {Prat}, {Prod'homme},
  {Raison}, {Rebordao}, {Risquez}, {Rocca-Volmerange}, {Rosen}, {Ruiz-Fuertes},
  {Russo}, {Sembay}, {Serraller Vizcaino}, {Short}, {Siebert}, {Silva},
  {Sinachopoulos}, {Slezak}, {Soffel}, {Sosnowska}, {Strai{\v{z}}ys}, {ter
  Linden}, {Terrell}, {Theil}, {Tiede}, {Troisi}, {Tsalmantza}, {Tur},
  {Vaccari}, {Vachier}, {Valles}, {Van Hamme}, {Veltz}, {Virtanen}, {Wallut},
  {Wichmann}, {Wilkinson}, {Ziaeepour}, \& {Zschocke}}]{Gaia_General_2016}
{Gaia Collaboration}, {Prusti}, T., {de Bruijne}, J.~H.~J., {et~al.} 2016,
  \aap, 595, A1, \dodoi{10.1051/0004-6361/201629272}

\bibitem[{{Gaia Collaboration} {et~al.}(2018{\natexlab{a}}){Gaia
  Collaboration}, {Brown}, {Vallenari}, {Prusti}, {de Bruijne}, {Babusiaux},
  {Bailer-Jones}, {Biermann}, {Evans}, {Eyer}, {Jansen}, {Jordi}, {Klioner},
  {Lammers}, {Lindegren}, {Luri}, {Mignard}, {Panem}, {Pourbaix}, {Randich},
  {Sartoretti}, {Siddiqui}, {Soubiran}, {van Leeuwen}, {Walton}, {Arenou},
  {Bastian}, {Cropper}, {Drimmel}, {Katz}, {Lattanzi}, {Bakker}, {Cacciari},
  {Casta{\~n}eda}, {Chaoul}, {Cheek}, {De Angeli}, {Fabricius}, {Guerra},
  {Holl}, {Masana}, {Messineo}, {Mowlavi}, {Nienartowicz}, {Panuzzo},
  {Portell}, {Riello}, {Seabroke}, {Tanga}, {Th{\'e}venin}, {Gracia-Abril},
  {Comoretto}, {Garcia-Reinaldos}, {Teyssier}, {Altmann}, {Andrae}, {Audard},
  {Bellas-Velidis}, {Benson}, {Berthier}, {Blomme}, {Burgess}, {Busso},
  {Carry}, {Cellino}, {Clementini}, {Clotet}, {Creevey}, {Davidson}, {De
  Ridder}, {Delchambre}, {Dell'Oro}, {Ducourant}, {Fern{\'a}ndez-
  Hern{\'a}ndez}, {Fouesneau}, {Fr{\'e}mat}, {Galluccio}, {Garc{\'\i}a-Torres},
  {Gonz{\'a}lez-N{\'u}{\~n}ez}, {Gonz{\'a}lez-Vidal}, {Gosset}, {Guy},
  {Halbwachs}, {Hambly}, {Harrison}, {Hern{\'a}ndez}, {Hestroffer}, {Hodgkin},
  {Hutton}, {Jasniewicz}, {Jean-Antoine-Piccolo}, {Jordan}, {Korn},
  {Krone-Martins}, {Lanzafame}, {Lebzelter}, {L{\"o}ffler}, {Manteiga},
  {Marrese}, {Mart{\'\i}n-Fleitas}, {Moitinho}, {Mora}, {Muinonen}, {Osinde},
  {Pancino}, {Pauwels}, {Petit}, {Recio-Blanco}, {Richards}, {Rimoldini},
  {Robin}, {Sarro}, {Siopis}, {Smith}, {Sozzetti}, {S{\"u}veges}, {Torra}, {van
  Reeven}, {Abbas}, {Abreu Aramburu}, {Accart}, {Aerts}, {Altavilla},
  {{\'A}lvarez}, {Alvarez}, {Alves}, {Anderson}, {Andrei}, {Anglada Varela},
  {Antiche}, {Antoja}, {Arcay}, {Astraatmadja}, {Bach}, {Baker},
  {Balaguer-N{\'u}{\~n}ez}, {Balm}, {Barache}, {Barata}, {Barbato}, {Barblan},
  {Barklem}, {Barrado}, {Barros}, {Barstow}, {Bartholom{\'e} Mu{\~n}oz},
  {Bassilana}, {Becciani}, {Bellazzini}, {Berihuete}, {Bertone}, {Bianchi},
  {Bienaym{\'e}}, {Blanco-Cuaresma}, {Boch}, {Boeche}, {Bombrun}, {Borrachero},
  {Bossini}, {Bouquillon}, {Bourda}, {Bragaglia}, {Bramante}, {Breddels},
  {Bressan}, {Brouillet}, {Br{\"u}semeister}, {Brugaletta}, {Bucciarelli},
  {Burlacu}, {Busonero}, {Butkevich}, {Buzzi}, {Caffau}, {Cancelliere},
  {Cannizzaro}, {Cantat-Gaudin}, {Carballo}, {Carlucci}, {Carrasco},
  {Casamiquela}, {Castellani}, {Castro-Ginard}, {Charlot}, {Chemin},
  {Chiavassa}, {Cocozza}, {Costigan}, {Cowell}, {Crifo}, {Crosta}, {Crowley},
  {Cuypers}, {Dafonte}, {Damerdji}, {Dapergolas}, {David}, {David}, {de
  Laverny}, {De Luise}, {De March}, {de Martino}, {de Souza}, {de Torres},
  {Debosscher}, {del Pozo}, {Delbo}, {Delgado}, {Delgado}, {Di Matteo},
  {Diakite}, {Diener}, {Distefano}, {Dolding}, {Drazinos}, {Dur{\'a}n},
  {Edvardsson}, {Enke}, {Eriksson}, {Esquej}, {Eynard Bontemps}, {Fabre},
  {Fabrizio}, {Faigler}, {Falc{\~a}o}, {Farr{\`a}s Casas}, {Federici},
  {Fedorets}, {Fernique}, {Figueras}, {Filippi}, {Findeisen}, {Fonti},
  {Fraile}, {Fraser}, {Fr{\'e}zouls}, {Gai}, {Galleti}, {Garabato},
  {Garc{\'\i}a-Sedano}, {Garofalo}, {Garralda}, {Gavel}, {Gavras}, {Gerssen},
  {Geyer}, {Giacobbe}, {Gilmore}, {Girona}, {Giuffrida}, {Glass}, {Gomes},
  {Granvik}, {Gueguen}, {Guerrier}, {Guiraud}, {Guti{\'e}rrez-S{\'a}nchez},
  {Haigron}, {Hatzidimitriou}, {Hauser}, {Haywood}, {Heiter}, {Helmi}, {Heu},
  {Hilger}, {Hobbs}, {Hofmann}, {Holland}, {Huckle}, {Hypki}, {Icardi},
  {Jan{\ss}en}, {Jevardat de Fombelle}, {Jonker}, {Juh{\'a}sz}, {Julbe},
  {Karampelas}, {Kewley}, {Klar}, {Kochoska}, {Kohley}, {Kolenberg},
  {Kontizas}, {Kontizas}, {Koposov}, {Kordopatis}, {Kostrzewa-Rutkowska},
  {Koubsky}, {Lambert}, {Lanza}, {Lasne}, {Lavigne}, {Le Fustec}, {Le
  Poncin-Lafitte}, {Lebreton}, {Leccia}, {Leclerc}, {Lecoeur-Taibi},
  {Lenhardt}, {Leroux}, {Liao}, {Licata}, {Lindstr{\o}m}, {Lister}, {Livanou},
  {Lobel}, {L{\'o}pez}, {Managau}, {Mann}, {Mantelet}, {Marchal}, {Marchant},
  {Marconi}, {Marinoni}, {Marschalk{\'o}}, {Marshall}, {Martino}, {Marton},
  {Mary}, {Massari}, {Matijevi{\v{c}}}, {Mazeh}, {McMillan}, {Messina},
  {Michalik}, {Millar}, {Molina}, {Molinaro}, {Moln{\'a}r}, {Montegriffo},
  {Mor}, {Morbidelli}, {Morel}, {Morris}, {Mulone}, {Muraveva}, {Musella},
  {Nelemans}, {Nicastro}, {Noval}, {O'Mullane}, {Ord{\'e}novic},
  {Ord{\'o}{\~n}ez-Blanco}, {Osborne}, {Pagani}, {Pagano}, {Pailler},
  {Palacin}, {Palaversa}, {Panahi}, {Pawlak}, {Piersimoni}, {Pineau}, {Plachy},
  {Plum}, {Poggio}, {Poujoulet}, {Pr{\v{s}}a}, {Pulone}, {Racero}, {Ragaini},
  {Rambaux}, {Ramos-Lerate}, {Regibo}, {Reyl{\'e}}, {Riclet}, {Ripepi}, {Riva},
  {Rivard}, {Rixon}, {Roegiers}, {Roelens}, {Romero-G{\'o}mez}, {Rowell},
  {Royer}, {Ruiz-Dern}, {Sadowski}, {Sagrist{\`a} Sell{\'e}s}, {Sahlmann},
  {Salgado}, {Salguero}, {Sanna}, {Santana- Ros}, {Sarasso}, {Savietto},
  {Schultheis}, {Sciacca}, {Segol}, {Segovia}, {S{\'e}gransan}, {Shih},
  {Siltala}, {Silva}, {Smart}, {Smith}, {Solano}, {Solitro}, {Sordo}, {Soria
  Nieto}, {Souchay}, {Spagna}, {Spoto}, {Stampa}, {Steele},
  {Steidelm{\"u}ller}, {Stephenson}, {Stoev}, {Suess}, {Surdej}, {Szabados},
  {Szegedi-Elek}, {Tapiador}, {Taris}, {Tauran}, {Taylor}, {Teixeira},
  {Terrett}, {Teyssandier}, {Thuillot}, {Titarenko}, {Torra Clotet}, {Turon},
  {Ulla}, {Utrilla}, {Uzzi}, {Vaillant}, {Valentini}, {Valette}, {van Elteren},
  {Van Hemelryck}, {van Leeuwen}, {Vaschetto}, {Vecchiato}, {Veljanoski},
  {Viala}, {Vicente}, {Vogt}, {von Essen}, {Voss}, {Votruba}, {Voutsinas},
  {Walmsley}, {Weiler}, {Wertz}, {Wevers}, {Wyrzykowski}, {Yoldas},
  {{\v{Z}}erjal}, {Ziaeepour}, {Zorec}, {Zschocke}, {Zucker}, {Zurbach}, \&
  {Zwitter}}]{Gaia_General_2018}
{Gaia Collaboration}, {Brown}, A.~G.~A., {Vallenari}, A., {et~al.}
  2018{\natexlab{a}}, \aap, 616, A1, \dodoi{10.1051/0004-6361/201833051}

\bibitem[{{Gaia Collaboration} {et~al.}(2018{\natexlab{b}}){Gaia
  Collaboration}, {Spoto}, {Tanga}, {Mignard}, {Berthier}, {Carry}, {Cellino},
  {Dell'Oro}, {Hestroffer}, {Muinonen}, {Pauwels}, {Petit}, {David}, {De
  Angeli}, {Delbo}, {Fr{\'e}zouls}, {Galluccio}, {Granvik}, {Guiraud},
  {Hern{\'a}ndez}, {Ord{\'e}novic}, {Portell}, {Poujoulet}, {Thuillot},
  {Walmsley}, {Brown}, {Vallenari}, {Prusti}, {de Bruijne}, {Babusiaux},
  {Bailer-Jones}, {Biermann}, {Evans}, {Eyer}, {Jansen}, {Jordi}, {Klioner},
  {Lammers}, {Lindegren}, {Luri}, {Panem}, {Pourbaix}, {Randich}, {Sartoretti},
  {Siddiqui}, {Soubiran}, {van Leeuwen}, {Walton}, {Arenou}, {Bastian},
  {Cropper}, {Drimmel}, {Katz}, {Lattanzi}, {Bakker}, {Cacciari},
  {Casta{\~n}eda}, {Chaoul}, {Cheek}, {Fabricius}, {Guerra}, {Holl}, {Masana},
  {Messineo}, {Mowlavi}, {Nienartowicz}, {Panuzzo}, {Riello}, {Seabroke},
  {Th{\'e}venin}, {Gracia-Abril}, {Comoretto}, {Garcia-Reinaldos}, {Teyssier},
  {Altmann}, {Andrae}, {Audard}, {Bellas-Velidis}, {Benson}, {Blomme},
  {Burgess}, {Busso}, {Clementini}, {Clotet}, {Creevey}, {Davidson}, {De
  Ridder}, {Delchambre}, {Ducourant}, {Fern{\'a}ndez-Hern{\'a}ndez},
  {Fouesneau}, {Fr{\'e}mat}, {Garc{\'\i}a-Torres},
  {Gonz{\'a}lez-N{\'u}{\~n}ez}, {Gonz{\'a}lez-Vidal}, {Gosset}, {Guy},
  {Halbwachs}, {Hambly}, {Harrison}, {Hodgkin}, {Hutton}, {Jasniewicz},
  {Jean-Antoine-Piccolo}, {Jordan}, {Korn}, {Krone-Martins}, {Lanzafame},
  {Lebzelter}, {L{\"o}}, {Manteiga}, {Marrese}, {Mart{\'\i}n-Fleitas},
  {Moitinho}, {Mora}, {Osinde}, {Pancino}, {Recio-Blanco}, {Richards},
  {Rimoldini}, {Robin}, {Sarro}, {Siopis}, {Smith}, {Sozzetti}, {S{\"u}veges},
  {Torra}, {van Reeven}, {Abbas}, {Abreu Aramburu}, {Accart}, {Aerts},
  {Altavilla}, {{\'A}lvarez}, {Alvarez}, {Alves}, {Anderson}, {Andrei},
  {Anglada Varela}, {Antiche}, {Antoja}, {Arcay}, {Astraatmadja}, {Bach},
  {Baker}, {Balaguer-N{\'u}{\~n}ez}, {Balm}, {Barache}, {Barata}, {Barbato},
  {Barblan}, {Barklem}, {Barrado}, {Barros}, {Barstow}, {Bartholom{\'e}
  Mu{\~n}oz}, {Bassilana}, {Becciani}, {Bellazzini}, {Berihuete}, {Bertone},
  {Bianchi}, {Bienaym{\'e}}, {Blanco-Cuaresma}, {Boch}, {Boeche}, {Bombrun},
  {Borrachero}, {Bossini}, {Bouquillon}, {Bourda}, {Bragaglia}, {Bramante},
  {Breddels}, {Bressan}, {Brouillet}, {Br{\"u}semeister}, {Brugaletta},
  {Bucciarelli}, {Burlacu}, {Busonero}, {Butkevich}, {Buzzi}, {Caffau},
  {Cancelliere}, {Cannizzaro}, {Cantat-Gaudin}, {Carballo}, {Carlucci},
  {Carrasco}, {Casamiquela}, {Castellani}, {Castro-Ginard}, {Charlot},
  {Chemin}, {Chiavassa}, {Cocozza}, {Costigan}, {Cowell}, {Crifo}, {Crosta},
  {Crowley}, {Cuypers}, {Dafonte}, {Damerdji}, {Dapergolas}, {David}, {de
  Laverny}, {De Luise}, {De March}, {de Souza}, {de Torres}, {Debosscher}, {del
  Pozo}, {Delgado}, {Delgado}, {Diakite}, {Diener}, {Distefano}, {Dolding},
  {Drazinos}, {Dur{\'a}n}, {Edvardsson}, {Enke}, {Eriksson}, {Esquej}, {Eynard
  Bontemps}, {Fabre}, {Fabrizio}, {Faigler}, {Falc{\~a}o}, {Farr{\`a}s Casas},
  {Federici}, {Fedorets}, {Fernique}, {Figueras}, {Filippi}, {Findeisen},
  {Fonti}, {Fraile}, {Fraser}, {Gai}, {Galleti}, {Garabato},
  {Garc{\'\i}a-Sedano}, {Garofalo}, {Garralda}, {Gavel}, {Gavras}, {Gerssen},
  {Geyer}, {Giacobbe}, {Gilmore}, {Girona}, {Giuffrida}, {Glass}, {Gomes},
  {Gueguen}, {Guerrier}, {Guti{\'e}}, {Haigron}, {Hatzidimitriou}, {Hauser},
  {Haywood}, {Heiter}, {Helmi}, {Heu}, {Hilger}, {Hobbs}, {Hofmann}, {Holland
  }, {Huckle}, {Hypki}, {Icardi}, {Jan{\ss}en}, {Jevardat de Fombelle},
  {Jonker}, {Juh{\'a}sz}, {Julbe}, {Karampelas}, {Kewley}, {Klar}, {Kochoska},
  {Kohley}, {Kolenberg}, {Kontizas}, {Kontizas}, {Koposov}, {Kordopatis},
  {Kostrzewa-Rutkowska}, {Koubsky}, {Lambert}, {Lanza}, {Lasne}, {Lavigne}, {Le
  Fustec}, {Le Poncin-Lafitte}, {Lebreton}, {Leccia}, {Leclerc},
  {Lecoeur-Taibi}, {Lenhardt}, {Leroux}, {Liao}, {Licata}, {Lindstr{\o}m},
  {Lister}, {Livanou}, {Lobel}, {L{\'o}pez}, {Managau}, {Mann}, {Mantelet},
  {Marchal}, {Marchant}, {Marconi}, {Marinoni}, {Marschalk{\'o}}, {Marshall},
  {Martino}, {Marton}, {Mary}, {Massari}, {Matijevi{\v{c}}}, {Mazeh},
  {McMillan}, {Messina}, {Michalik}, {Millar}, {Molina}, {Molinaro},
  {Moln{\'a}r}, {Montegriffo}, {Mor}, {Morbidelli}, {Morel}, {Morris},
  {Mulone}, {Muraveva}, {Musella}, {Nelemans}, {Nicastro}, {Noval},
  {O'Mullane}, {Ord{\'o}{\~n}ez-Blanco}, {Osborne}, {Pagani}, {Pagano},
  {Pailler}, {Palacin}, {Palaversa}, {Panahi}, {Pawlak}, {Piersimoni},
  {Pineau}, {Plachy}, {Plum}, {Poggio}, {Pr{\v{s}}a}, {Pulone}, {Racero},
  {Ragaini}, {Rambaux}, {Ramos-Lerate}, {Regibo}, {Reyl{\'e}}, {Riclet},
  {Ripepi}, {Riva}, {Rivard}, {Rixon}, {Roegiers}, {Roelens},
  {Romero-G{\'o}mez}, {Rowell}, {Royer}, {Ruiz-Dern}, {Sadowski}, {Sagrist{\`a}
  Sell{\'e}s}, {Sahlmann}, {Salgado}, {Salguero}, {Sanna}, {Santana-Ros},
  {Sarasso}, {Savietto}, {Schultheis}, {Sciacca}, {Segol}, {Segovia},
  {S{\'e}gransan}, {Shih}, {Siltala}, {Silva}, {Smart}, {Smith}, {Solano},
  {Solitro}, {Sordo}, {Soria Nieto}, {Souchay}, {Spagna}, {Stampa}, {Steele},
  {Steidelm{\"u}ller}, {Stephenson}, {Stoev}, {Suess}, {Surdej}, {Szabados},
  {Szegedi-Elek}, {Tapiador}, {Taris}, {Tauran}, {Taylor}, {Teixeira},
  {Terrett}, {Teyssand ier}, {Titarenko}, {Torra Clotet}, {Turon}, {Ulla},
  {Utrilla}, {Uzzi}, {Vaillant}, {Valentini}, {Valette}, {van Elteren}, {Van
  Hemelryck}, {van Leeuwen}, {Vaschetto}, {Vecchiato}, {Veljanoski}, {Viala},
  {Vicente}, {Vogt}, {von Essen}, {Voss}, {Votruba}, {Voutsinas}, {Weiler},
  {Wertz}, {Wevers}, {Wyrzykowski}, {Yoldas}, {{\v{Z}}erjal}, {Ziaeepour},
  {Zorec}, {Zschocke}, {Zucker}, {Zurbach}, \&
  {Zwitter}}]{Gaia_Solar_System_2018}
{Gaia Collaboration}, {Spoto}, F., {Tanga}, P., {et~al.} 2018{\natexlab{b}},
  \aap, 616, A13, \dodoi{10.1051/0004-6361/201832900}

\bibitem[{Gouda(2019)}]{gouda_2020}
Gouda, N. 2019, Proceedings of the International Astronomical Union, 14,
  51–53, \dodoi{10.1017/S1743921319007968}

\bibitem[{{Kervella} {et~al.}(2019){Kervella}, {Arenou}, {Mignard}, \&
  {Th{\'e}venin}}]{2019AA_Kervella}
{Kervella}, P., {Arenou}, F., {Mignard}, F., \& {Th{\'e}venin}, F. 2019, \aap,
  623, A72, \dodoi{10.1051/0004-6361/201834371}

\bibitem[{{Lagrange} {et~al.}(2020){Lagrange}, {Rubini}, {Nowak}, {Lacour},
  {Grandjean}, {Boccaletti}, {Langlois}, {Delorme}, {Gratton}, {Wang},
  {Flasseur}, {Galicher}, {Kral}, {Meunier}, {Beust}, {Babusiaux}, {Le
  Coroller}, {Thebault}, {Kervella}, {Zurlo}, {Maire}, {Wahhaj}, {Amorim},
  {Asensio-Torres}, {Benisty}, {Berger}, {Bonnefoy}, {Brandner}, {Cantalloube},
  {Charnay}, {Chauvin}, {Choquet}, {Cl{\'e}net}, {Christiaens}, {Coud{\'e} Du
  Foresto}, {de Zeeuw}, {Desidera}, {Duvert}, {Eckart}, {Eisenhauer},
  {Galland}, {Gao}, {Garcia}, {Garcia Lopez}, {Gendron}, {Genzel}, {Gillessen},
  {Girard}, {Hagelberg}, {Haubois}, {Henning}, {Heissel}, {Hippler},
  {Horrobin}, {Janson}, {Kammerer}, {Kenworthy}, {Keppler}, {Kreidberg},
  {Lapeyr{\`e}re}, {Le Bouquin}, {L{\'e}na}, {M{\'e}rand}, {Messina},
  {Molli{\`e}re}, {Monnier}, {Ott}, {Otten}, {Paumard}, {Paladini}, {Perraut},
  {Perrin}, {Pueyo}, {Pfuhl}, {Rodet}, {Rodriguez-Coira}, {Rousset}, {Samland
  }, {Shangguan}, {Schmidt}, {Straub}, {Straubmeier}, {Stolker}, {Vigan},
  {Vincent}, {Widmann}, {Woillez}, \& {Gravity
  Collaboration}}]{AMLagrange2020betapicc_direct_detection}
{Lagrange}, A.~M., {Rubini}, P., {Nowak}, M., {et~al.} 2020, \aap, 642, A18,
  \dodoi{10.1051/0004-6361/202038823}

\bibitem[{{Li} {et~al.}(2021){Li}, {Brandt}, {Brandt}, {Dupuy}, {Michalik},
  {Jensen-Clem}, {Zeng}, {Faherty}, \&
  {Mitra}}]{2021_Li_Yiting_nine_masses_RV_planets}
{Li}, Y., {Brandt}, T.~D., {Brandt}, G.~M., {et~al.} 2021, arXiv e-prints,
  arXiv:2109.10422.
\newblock \doarXiv{2109.10422}

\bibitem[{{Lindegren} \& {Dravins}(2021)}]{2021arXiv210509014L}
{Lindegren}, L., \& {Dravins}, D. 2021, arXiv e-prints, arXiv:2105.09014.
\newblock \doarXiv{2105.09014}

\bibitem[{{Lindegren} {et~al.}(1997){Lindegren}, {Mignard}, {S{\"o}derhjelm},
  {Badiali}, {Bernstein}, {Lampens}, {Pannunzio}, {Arenou}, {Bernacca},
  {Falin}, {Froeschl{\'e}}, {Kovalevsky}, {Martin}, {Perryman}, \&
  {Wielen}}]{1997A&A...323L..53L}
{Lindegren}, L., {Mignard}, F., {S{\"o}derhjelm}, S., {et~al.} 1997, \aap, 323,
  L53

\bibitem[{{Lindegren} {et~al.}(2018){Lindegren}, {Hernandez}, {Bombrun},
  {Klioner}, {Bastian}, {Ramos-Lerate}, {de Torres}, {Steidelmuller},
  {Stephenson}, {Hobbs}, {Lammers}, {Biermann}, {Geyer}, {Hilger}, {Michalik},
  {Stampa}, {McMillan}, {Castaneda}, {Clotet}, {Comoretto}, {Davidson},
  {Fabricius}, {Gracia}, {Hambly}, {Hutton}, {Mora}, {Portell}, {van Leeuwen},
  {Abbas}, {Abreu}, {Altmann}, {Andrei}, {Anglada}, {Balaguer-Nunez},
  {Barache}, {Becciani}, {Bertone}, {Bianchi}, {Bouquillon}, {Bourda},
  {Brusemeister}, {Bucciarelli}, {Busonero}, {Buzzi}, {Cancelliere},
  {Carlucci}, {Charlot}, {Cheek}, {Crosta}, {Crowley}, {de Bruijne}, {de
  Felice}, {Drimmel}, {Esquej}, {Fienga}, {Fraile}, {Gai}, {Garralda},
  {Gonzalez-Vidal}, {Guerra}, {Hauser}, {Hofmann}, {Holl}, {Jordan},
  {Lattanzi}, {Lenhardt}, {Liao}, {Licata}, {Lister}, {Loffler}, {Marchant},
  {Martin-Fleitas}, {Messineo}, {Mignard}, {Morbidelli}, {Poggio}, {Riva},
  {Rowell}, {Salguero}, {Sarasso}, {Sciacca}, {Siddiqui}, {Smart}, {Spagna},
  {Steele}, {Taris}, {Torra}, {van Elteren}, {van Reeven}, \&
  {Vecchiato}}]{Gaia_Astrometry_2018}
{Lindegren}, L., {Hernandez}, J., {Bombrun}, A., {et~al.} 2018, arxiv.
\newblock \doarXiv{1804.09366}

\bibitem[{{Lindegren} {et~al.}(2020){Lindegren}, {Klioner}, {Hern{\'a}ndez},
  {Bombrun}, {Ramos-Lerate}, {Steidelm{\"u}ller}, {Bastian}, {Biermann}, {de
  Torres}, {Gerlach}, {Geyer}, {Hilger}, {Hobbs}, {Lammers}, {McMillan},
  {Stephenson}, {Casta{\~n}eda}, {Davidson}, {Fabricius}, {Gracia-Abril},
  {Portell}, {Rowell}, {Teyssier}, {Torra}, {Bartolom{\'e}}, {Clotet},
  {Garralda}, {Gonz{\'a}lez-Vidal}, {Torra}, {Abbas}, {Altmann}, {Anglada
  Varela}, {Balaguer-N{\'u}{\~n}ez}, {Balog}, {Barache}, {Becciani}, {Bernet},
  {Bertone}, {Bianchi}, {Bouquillon}, {Brown}, {Bucciarelli}, {Busonero},
  {Butkevich}, {Buzzi}, {Cancelliere}, {Carlucci}, {Charlot}, {Cioni},
  {Crosta}, {Crowley}, {del Peloso}, {del Pozo}, {Drimmel}, {Esquej}, {Fienga},
  {Fraile}, {Gai}, {Garcia-Reinaldos}, {Guerra}, {Hambly}, {Hauser},
  {Jan{\ss}en}, {Jordan}, {Kostrzewa-Rutkowska}, {Lattanzi}, {Liao}, {Licata},
  {Lister}, {L{\"o}ffler}, {Marchant}, {Masip}, {Mignard}, {Mints}, {Molina},
  {Mora}, {Morbidelli}, {Murphy}, {Pagani}, {Panuzzo}, {Pe{\~n}alosa Esteller},
  {Poggio}, {Re Fiorentin}, {Riva}, {Sagrist{\`a} Sell{\'e}s}, {Sanchez
  Gimenez}, {Sarasso}, {Sciacca}, {Siddiqui}, {Smart}, {Souami}, {Spagna},
  {Steele}, {Taris}, {Utrilla}, {van Reeven}, \&
  {Vecchiato}}]{Lindegren+Klioner+Hernandez+etal_2020}
{Lindegren}, L., {Klioner}, S.~A., {Hern{\'a}ndez}, J., {et~al.} 2020, arXiv
  e-prints, arXiv:2012.03380.
\newblock \doarXiv{2012.03380}

\bibitem[{{Luyten}(1979)}]{Luyten-high-proper}
{Luyten}, W.~J. 1979, {LHS catalogue. A catalogue of stars with proper motions
  exceeding 0''5 annually}

\bibitem[{{Maire} {et~al.}(2020){Maire}, {Molaverdikhani}, {Desidera},
  {Trifonov}, {Molli{\`e}re}, {D'Orazi}, {Frankel}, {Baudino}, {Messina},
  {M{\"u}ller}, {Charnay}, {Cheetham}, {Delorme}, {Ligi}, {Bonnefoy},
  {Brandner}, {Mesa}, {Cantalloube}, {Galicher}, {Henning}, {Biller},
  {Hagelberg}, {Lagrange}, {Lavie}, {Rickman}, {S{\'e}gransan}, {Udry},
  {Chauvin}, {Gratton}, {Langlois}, {Vigan}, {Meyer}, {Beuzit}, {Bhowmik},
  {Boccaletti}, {Lazzoni}, {Perrot}, {Schmidt}, {Zurlo}, {Gluck}, {Pragt},
  {Ramos}, {Roelfsema}, {Roux}, \& {Sauvage}}]{2020AA_Maire_HD19467}
{Maire}, A.~L., {Molaverdikhani}, K., {Desidera}, S., {et~al.} 2020, \aap, 639,
  A47, \dodoi{10.1051/0004-6361/202037984}

\bibitem[{{Malbet} {et~al.}(2019){Malbet}, {Abbas}, {Alves}, {Boehm}, {Brown},
  {Chemin}, {Correia}, {Courbin}, {Darling}, {Diaferio}, {Fortin}, {Fridlund},
  {Gnedin}, {Holl}, {Krone-Martins}, {L{\'e}ger}, {Labadie}, {Laskar}, {Mamon},
  {McArthur}, {Michalik}, {Moitinho}, {Oertel}, {Ostorero}, {Schneider},
  {Scott}, {Shao}, {Sozzetti}, {Tomsick}, {Valluri}, \&
  {Wyse}}]{2019arXiv191008028M}
{Malbet}, F., {Abbas}, U., {Alves}, J., {et~al.} 2019, arXiv e-prints,
  arXiv:1910.08028.
\newblock \doarXiv{1910.08028}

\bibitem[{{Melchior} {et~al.}(2018){Melchior}, {Spergel}, \&
  {Lanz}}]{Melchor+Spergel+Lanz_2018}
{Melchior}, P., {Spergel}, D., \& {Lanz}, A. 2018, \aj, 155, 102,
  \dodoi{10.3847/1538-3881/aaa422}

\bibitem[{{Michalik} {et~al.}(2014){Michalik}, {Lindegren}, {Hobbs}, \&
  {Lammers}}]{Michalik+Lindegren+Hobbs+etal_2014}
{Michalik}, D., {Lindegren}, L., {Hobbs}, D., \& {Lammers}, U. 2014, \aap, 571,
  A85, \dodoi{10.1051/0004-6361/201424606}

\bibitem[{{Nielsen} {et~al.}(2019){Nielsen}, {De Rosa}, {Wang}, {Sahlmann},
  {Kalas}, {Duchene}, {Rameau}, {Marley}, {Saumon}, {Macintosh},
  {Millar-Blanchaer}, {Nguyen}, {Ammons}, {Bailey}, {Barman}, {Bulger},
  {Chilcote}, {Cotten}, {Doyon}, {Esposito}, {Fitzgerald}, {Follette},
  {Gerard}, {Goodsell}, {Graham}, {Greenbaum}, {Hibon}, {Hung}, {Ingraham},
  {Konopacky}, {Larkin}, {Maire}, {Marchis}, {Marois}, {Metchev},
  {Oppenheimer}, {Palmer}, {Patience}, {Perrin}, {Poyneer}, {Pueyo}, {Rajan},
  {Rantakyro}, {Ruffio}, {Savransky}, {Schneider}, {Sivaramakrishnan}, {Song},
  {Soummer}, {Thomas}, {Wallace}, {Ward-Duong}, {Wiktorowicz}, \&
  {Wolff}}]{Nielson+DeRosa+etal+betapicc2019}
{Nielsen}, E.~L., {De Rosa}, R.~J., {Wang}, J.~J., {et~al.} 2019, arXiv
  e-prints, arXiv:1911.11273.
\newblock \doarXiv{1911.11273}

\bibitem[{Oliphant(2006)}]{numpy1}
Oliphant, T. 2006, {NumPy}: A guide to {NumPy}, USA: Trelgol Publishing.
\newblock \url{http://www.numpy.org/}

\bibitem[{pandas~development team(2020)}]{reback2020pandas}
pandas~development team, T. 2020, pandas-dev/pandas: Pandas, 1.0.5,  Zenodo,
  \dodoi{10.5281/zenodo.3509134}

\bibitem[{{Perryman}(2012)}]{2012PerrymanAstrometryHistory}
{Perryman}, M. 2012, European Physical Journal H, 37, 745,
  \dodoi{10.1140/epjh/e2012-30039-4}

\bibitem[{{Perryman} {et~al.}(2014){Perryman}, {Hartman}, {Bakos}, \&
  {Lindegren}}]{Perryman+Hartman+Bakos+etal_2014}
{Perryman}, M., {Hartman}, J., {Bakos}, G.~{\'A}., \& {Lindegren}, L. 2014,
  \apj, 797, 14, \dodoi{10.1088/0004-637X/797/1/14}

\bibitem[{{Perryman} {et~al.}(1997){Perryman}, {Lindegren}, {Kovalevsky},
  {Hog}, {Bastian}, {Bernacca}, {Creze}, {Donati}, {Grenon}, {Grewing}, {van
  Leeuwen}, {van der Marel}, {Mignard}, {Murray}, {Le Poole}, {Schrijver},
  {Turon}, {Arenou}, {Froeschle}, \& {Petersen}}]{HIP_TYCHO_ESA_1997}
{Perryman}, M.~A.~C., {Lindegren}, L., {Kovalevsky}, J., {et~al.} 1997, \aap,
  500, 501

\bibitem[{{Price-Whelan} {et~al.}(2018){Price-Whelan}, {Sip{\H{o}}cz},
  {G{\"u}nther}, {Lim}, {Crawford}, {Conseil}, {Shupe}, {Craig}, {Dencheva},
  {Ginsburg}, {VanderPlas}, {Bradley}, {P{\'e}rez-Su{\'a}rez}, {de Val-Borro},
  {Paper Contributors}, {Aldcroft}, {Cruz}, {Robitaille}, {Tollerud},
  {Coordination Committee}, {Ardelean}, {Babej}, {Bach}, {Bachetti}, {Bakanov},
  {Bamford}, {Barentsen}, {Barmby}, {Baumbach}, {Berry}, {Biscani}, {Boquien},
  {Bostroem}, {Bouma}, {Brammer}, {Bray}, {Breytenbach}, {Buddelmeijer},
  {Burke}, {Calderone}, {Cano Rodr{\'\i}guez}, {Cara}, {Cardoso}, {Cheedella},
  {Copin}, {Corrales}, {Crichton}, {D{\textquoteright}Avella}, {Deil},
  {Depagne}, {Dietrich}, {Donath}, {Droettboom}, {Earl}, {Erben}, {Fabbro},
  {Ferreira}, {Finethy}, {Fox}, {Garrison}, {Gibbons}, {Goldstein}, {Gommers},
  {Greco}, {Greenfield}, {Groener}, {Grollier}, {Hagen}, {Hirst}, {Homeier},
  {Horton}, {Hosseinzadeh}, {Hu}, {Hunkeler}, {Ivezi{\'c}}, {Jain}, {Jenness},
  {Kanarek}, {Kendrew}, {Kern}, {Kerzendorf}, {Khvalko}, {King}, {Kirkby},
  {Kulkarni}, {Kumar}, {Lee}, {Lenz}, {Littlefair}, {Ma}, {Macleod},
  {Mastropietro}, {McCully}, {Montagnac}, {Morris}, {Mueller}, {Mumford},
  {Muna}, {Murphy}, {Nelson}, {Nguyen}, {Ninan}, {N{\"o}the}, {Ogaz}, {Oh},
  {Parejko}, {Parley}, {Pascual}, {Patil}, {Patil}, {Plunkett}, {Prochaska},
  {Rastogi}, {Reddy Janga}, {Sabater}, {Sakurikar}, {Seifert}, {Sherbert},
  {Sherwood-Taylor}, {Shih}, {Sick}, {Silbiger}, {Singanamalla}, {Singer},
  {Sladen}, {Sooley}, {Sornarajah}, {Streicher}, {Teuben}, {Thomas},
  {Tremblay}, {Turner}, {Terr{\'o}n}, {van Kerkwijk}, {de la Vega}, {Watkins},
  {Weaver}, {Whitmore}, {Woillez}, {Zabalza}, \& {Contributors}}]{astropy:2018}
{Price-Whelan}, A.~M., {Sip{\H{o}}cz}, B.~M., {G{\"u}nther}, H.~M., {et~al.}
  2018, \aj, 156, 123, \dodoi{10.3847/1538-3881/aabc4f}

\bibitem[{{Quirrenbach} {et~al.}(2011){Quirrenbach}, {Reffert}, \&
  {Bergmann}}]{2011AIPC.1331..102Q}
{Quirrenbach}, A., {Reffert}, S., \& {Bergmann}, C. 2011, in American Institute
  of Physics Conference Series, Vol. 1331, Planetary Systems Beyond the Main
  Sequence, ed. S.~{Schuh}, H.~{Drechsel}, \& U.~{Heber}, 102--109,
  \dodoi{10.1063/1.3556189}

\bibitem[{{Reffert} \& {Quirrenbach}(2011)}]{2011A&A...527A.140R}
{Reffert}, S., \& {Quirrenbach}, A. 2011, \aap, 527, A140,
  \dodoi{10.1051/0004-6361/201015861}

\bibitem[{{Sahlmann} {et~al.}(2011){Sahlmann}, {S{\'e}gransan}, {Queloz},
  {Udry}, {Santos}, {Marmier}, {Mayor}, {Naef}, {Pepe}, \&
  {Zucker}}]{Sahlmann+Segransan+Queloz_2011}
{Sahlmann}, J., {S{\'e}gransan}, D., {Queloz}, D., {et~al.} 2011, \aap, 525,
  A95, \dodoi{10.1051/0004-6361/201015427}

\bibitem[{{Sanderson} {et~al.}(2017){Sanderson}, {Bellini}, {Casertano}, {Lu},
  {Melchior}, {Libralato}, {Bennett}, {Shao}, {Rhodes}, {Sohn}, {Malhotra},
  {Gaudi}, {Fall}, {Nelan}, {Guhathakurta}, {Anderson}, \&
  {Ho}}]{Sanderson+Bellini+etal_WFIRSTprecision}
{Sanderson}, R.~E., {Bellini}, A., {Casertano}, S., {et~al.} 2017, arXiv
  e-prints, arXiv:1712.05420.
\newblock \doarXiv{1712.05420}

\bibitem[{{Shao} {et~al.}(2009){Shao}, {Marcy}, {Catanzarite}, {Edberg},
  {L{\'e}ger}, {Malbet}, {Queloz}, {Muterspaugh}, {Beichman}, {Fischer},
  {Ford}, {Olling}, {Kulkarni}, {Unwin}, \& {Traub}}]{2009_SimLITE}
{Shao}, M., {Marcy}, G., {Catanzarite}, J.~H., {et~al.} 2009, in astro2010: The
  Astronomy and Astrophysics Decadal Survey, Vol. 2010, 271.
\newblock \doarXiv{0904.0965}

\bibitem[{{Snellen} \& {Brown}(2018)}]{Snellen+Brown_2018}
{Snellen}, I.~A.~G., \& {Brown}, A.~G.~A. 2018, Nature Astronomy, 2, 883,
  \dodoi{10.1038/s41550-018-0561-6}

\bibitem[{{Sozzetti} \& {Desidera}(2010)}]{2010A&A...509A.103S}
{Sozzetti}, A., \& {Desidera}, S. 2010, \aap, 509, A103,
  \dodoi{10.1051/0004-6361/200912717}

\bibitem[{{Spergel} {et~al.}(2015){Spergel}, {Gehrels}, {Baltay}, {Bennett},
  {Breckinridge}, {Donahue}, {Dressler}, {Gaudi}, {Greene}, {Guyon}, {Hirata},
  {Kalirai}, {Kasdin}, {Macintosh}, {Moos}, {Perlmutter}, {Postman},
  {Rauscher}, {Rhodes}, {Wang}, {Weinberg}, {Benford}, {Hudson}, {Jeong},
  {Mellier}, {Traub}, {Yamada}, {Capak}, {Colbert}, {Masters}, {Penny},
  {Savransky}, {Stern}, {Zimmerman}, {Barry}, {Bartusek}, {Carpenter}, {Cheng},
  {Content}, {Dekens}, {Demers}, {Grady}, {Jackson}, {Kuan}, {Kruk}, {Melton},
  {Nemati}, {Parvin}, {Poberezhskiy}, {Peddie}, {Ruffa}, {Wallace}, {Whipple},
  {Wollack}, \& {Zhao}}]{wfirst_spergel_2015}
{Spergel}, D., {Gehrels}, N., {Baltay}, C., {et~al.} 2015, arXiv e-prints,
  arXiv:1503.03757.
\newblock \doarXiv{1503.03757}

\bibitem[{{The Theia Collaboration} {et~al.}(2017){The Theia Collaboration},
  {Boehm}, {Krone-Martins}, {Amorim}, {Anglada-Escude}, {Brandeker}, {Courbin},
  {Ensslin}, {Falcao}, {Freese}, {Holl}, {Labadie}, {Leger}, {Malbet}, {Mamon},
  {McArthur}, {Mora}, {Shao}, {Sozzetti}, {Spolyar}, {Villaver}, {Albertus},
  {Bertone}, {Bouy}, {Boylan-Kolchin}, {Brown}, {Brown}, {Cardoso}, {Chemin},
  {Claudi}, {Correia}, {Crosta}, {Crouzier}, {Cyr-Racine}, {Damasso}, {da
  Silva}, {Davies}, {Das}, {Dayal}, {de Val-Borro}, {Diaferio}, {Erickcek},
  {Fairbairn}, {Fortin}, {Fridlund}, {Garcia}, {Gnedin}, {Goobar}, {Gordo},
  {Goullioud}, {Hambly}, {Hara}, {Hobbs}, {Hog}, {Holland}, {Ibata}, {Jordi},
  {Klioner}, {Kopeikin}, {Lacroix}, {Laskar}, {Le Poncin-Lafitte}, {Luri},
  {Majumdar}, {Makarov}, {Massey}, {Mennesson}, {Michalik}, {Moitinho de
  Almeida}, {Mourao}, {Moustakas}, {Murray}, {Muterspaugh}, {Oertel},
  {Ostorero}, {Perez-Garcia}, {Platais}, {de Mora}, {Quirrenbach}, {Randall},
  {Read}, {Regos}, {Rory}, {Rybicki}, {Scott}, {Schneider}, {Scholtz},
  {Siebert}, {Tereno}, {Tomsick}, {Traub}, {Valluri}, {Walker}, {Walton},
  {Watkins}, {White}, {Evans}, {Wyrzykowski}, \& {Wyse}}]{2017arXiv170701348T}
{The Theia Collaboration}, {Boehm}, C., {Krone-Martins}, A., {et~al.} 2017,
  arXiv e-prints, arXiv:1707.01348.
\newblock \doarXiv{1707.01348}

\bibitem[{{Unwin} {et~al.}(2008){Unwin}, {Shao}, {Tanner}, {Allen}, {Beichman},
  {Boboltz}, {Catanzarite}, {Chaboyer}, {Ciardi}, {Edberg}, {Fey}, {Fischer},
  {Gelino}, {Gould}, {Grillmair}, {Henry}, {Johnston}, {Johnston}, {Jones},
  {Kulkarni}, {Law}, {Majewski}, {Makarov}, {Marcy}, {Meier}, {Olling}, {Pan},
  {Patterson}, {Pitesky}, {Quirrenbach}, {Shaklan}, {Shaya}, {Strigari},
  {Tomsick}, {Wehrle}, \& {Worthey}}]{2008PASPSIM_PlanetQuest}
{Unwin}, S.~C., {Shao}, M., {Tanner}, A.~M., {et~al.} 2008, \pasp, 120, 38,
  \dodoi{10.1086/525059}

\bibitem[{{van de Kamp}(1935)}]{1935AJ.....44...74V}
{van de Kamp}, P. 1935, \aj, 44, 74, \dodoi{10.1086/105263}

\bibitem[{{van der Walt} {et~al.}(2011){van der Walt}, {Colbert}, \&
  {Varoquaux}}]{numpy2}
{van der Walt}, S., {Colbert}, S.~C., \& {Varoquaux}, G. 2011, Computing in
  Science and Engineering, 13, 22, \dodoi{10.1109/MCSE.2011.37}

\bibitem[{{van Leeuwen}(2007{\natexlab{a}})}]{vanLeeuwen_2007}
{van Leeuwen}, F. 2007{\natexlab{a}}, \aap, 474, 653,
  \dodoi{10.1051/0004-6361:20078357}

\bibitem[{{van Leeuwen}(2007{\natexlab{b}})}]{hipparcos_rereduction_book}
---. 2007{\natexlab{b}}, {Hipparcos, the New Reduction of the Raw Data}, Vol.
  350, \dodoi{10.1007/978-1-4020-6342-8}

\bibitem[{{van Leeuwen} \& {Evans}(1998)}]{1998AAS130157V}
{van Leeuwen}, F., \& {Evans}, D.~W. 1998, Astronomy and Astrophysics
  Supplement Series, 130, 157, \dodoi{10.1051/aas:1998218}

\bibitem[{{van Leeuwen} \& {Michalik}(2019)}]{floor_private_comm}
{van Leeuwen}, F., \& {Michalik}, D. 2019, Private Communication

\bibitem[{{van Leeuwen} \& {Michalik,
  D}(2021)}]{private_comm_floor_june_8_2021}
{van Leeuwen}, F., \& {Michalik, D}. 2021, Private Communication

\bibitem[{{Venner} {et~al.}(2021){Venner}, {Vanderburg}, \&
  {Pearce}}]{2021arXiv210413941V}
{Venner}, A., {Vanderburg}, A., \& {Pearce}, L.~A. 2021, arXiv e-prints,
  arXiv:2104.13941.
\newblock \doarXiv{2104.13941}

\bibitem[{{Virtanen} {et~al.}(2020){Virtanen}, {Gommers}, {Oliphant},
  {Haberland}, {Reddy}, {Cournapeau}, {Burovski}, {Peterson}, {Weckesser},
  {Bright}, {van der Walt}, {Brett}, {Wilson}, {Jarrod Millman}, {Mayorov},
  {Nelson}, {Jones}, {Kern}, {Larson}, {Carey}, {Polat}, {Feng}, {Moore}, {Vand
  erPlas}, {Laxalde}, {Perktold}, {Cimrman}, {Henriksen}, {Quintero}, {Harris},
  {Archibald}, {Ribeiro}, {Pedregosa}, {van Mulbregt}, \&
  {Contributors}}]{2020SciPy-NMeth}
{Virtanen}, P., {Gommers}, R., {Oliphant}, T.~E., {et~al.} 2020, Nature
  Methods, 17, 261, \dodoi{https://doi.org/10.1038/s41592-019-0686-2}

\bibitem[{{Vousden} {et~al.}(2016){Vousden}, {Farr}, \&
  {Mandel}}]{Vousden+Farr+Mandel_2016}
{Vousden}, W.~D., {Farr}, W.~M., \& {Mandel}, I. 2016, \mnras, 455, 1919,
  \dodoi{10.1093/mnras/stv2422}

\bibitem[{{W}es {M}c{K}inney(2010)}]{mckinney-proc-scipy-2010}
{W}es {M}c{K}inney. 2010, in {P}roceedings of the 9th {P}ython in {S}cience
  {C}onference, ed. {S}t\'efan van~der {W}alt \& {J}arrod {M}illman, 56 -- 61,
  \dodoi{10.25080/Majora-92bf1922-00a}

\bibitem[{{Zucker} \& {Mazeh}(2001)}]{2001ApJ_Zucker_Shay}
{Zucker}, S., \& {Mazeh}, T. 2001, \apj, 562, 549, \dodoi{10.1086/322959}

\end{thebibliography}
\bibliographystyle{aasjournal}

\end{document}